\documentclass[journal]{IEEEtran}
\usepackage{amsmath,amsfonts}
\usepackage{algorithmic}
\usepackage{algorithm}
\usepackage{array}
\usepackage[caption=false,font=normalsize,labelfont=sf,textfont=sf]{subfig}
\usepackage{textcomp}
\usepackage{stfloats}
\usepackage{url}
\usepackage{verbatim}
\usepackage{graphicx}

\usepackage{booktabs}
\usepackage[table,xcdraw]{xcolor}
\usepackage{hyperref}
 \hypersetup{
     colorlinks = true,
     allcolors = {black},
    }
\usepackage{cleveref}
\usepackage{array}
\usepackage{gensymb}
\usepackage{soul}
\usepackage[numbers]{natbib}

\begin{document}

\title{EnVisionVR: A Scene Interpretation Tool for Visual Accessibility in Virtual Reality}


\author{Junlong Chen\thanks{Email: jc2375@cam.ac.uk}, 
Rosella P. Galindo Esparza\thanks{Email: rosellapaulina.galindoesparza@brunel.ac.uk}, 
Vanja Garaj\thanks{Email: vanja.garaj@brunel.ac.uk}, 
Per Ola Kristensson\thanks{Email: pok21@cam.ac.uk}, 
John Dudley\thanks{Email: jjd50@cam.ac.uk}}

\maketitle

\begin{abstract}
Effective visual accessibility in Virtual Reality (VR) is crucial for Blind and Low Vision (BLV) users. However, designing visual accessibility systems is challenging due to the complexity of 3D VR environments and the need for techniques that can be easily retrofitted into existing applications. 
While prior work has studied how to enhance or translate visual information, the advancement of Vision Language Models (VLMs) provides an exciting opportunity to advance the scene interpretation capability of current systems. 
This paper presents \textsc{EnVisionVR}, an accessibility tool for VR scene interpretation. 
Through a formative study of usability barriers, we confirmed the lack of visual accessibility features as a key barrier for BLV users of VR content and applications.
In response, we designed and developed \textsc{EnVisionVR}, a novel visual accessibility system leveraging a VLM, voice input and multimodal feedback for scene interpretation and virtual object interaction in VR.
An evaluation with 12 BLV users demonstrated that \textsc{EnVisionVR} significantly improved their ability to locate virtual objects, effectively supporting scene understanding and object interaction.
\end{abstract}


\section{Introduction}
\IEEEPARstart{V}{irtual} Reality (VR) is a primarily visual medium.
The centrality of visual perception in the VR experience presents a major challenge when making the technology accessible to Blind and Low Vision (BLV) users. While screen readers and voiceover systems have played a crucial role in enabling BLV users to access information from two-dimensional (2D) screens, this accessibility issue persists 
for three-dimensional (3D) spatial content.
In contrast with how a screen reader works on conventional 2D user interfaces, the current form of VR applications challenges the systematic organisation and delivery of 3D spatial information in an intuitive and efficient format.

In an effort to address the exclusion of BLV users from VR experiences, prior work has studied visual accessibility design in virtual \cite{zhao2019seeingvr, canetroller} and augmented reality \cite{herskovitz2020making}. These efforts have adopted different strategies, such as enhancing visual information through view magnification, brightness/contrast adjustment, object contour highlighting~\cite{zhao2019seeingvr}; or converting visual information to other forms like audio descriptions of virtual objects~\cite{zhao2019seeingvr} or vibrotactile feedback~\cite{canetroller}.
With the advent of Vision Language Models (VLMs), new opportunities are emerging to generate vivid and detailed scene descriptions based on the user's field of view.
Such capability can be embedded in output modalities such as speech, audio, and haptic cues to facilitate the user's understanding of 3D scenes. 

This paper presents \textsc{EnVisionVR}, an integrated set of VR scene interpretation and virtual object localization tools that assist BLV users in navigating VR. 
The development was guided by a formative usability study with nine BLV participants, 
which provided empirical information on the support required by BLV users and, particularly, the types of accessibility features that support scene understanding and interaction. 
\textsc{EnVisionVR} was then implemented as a proof-of-concept system to improve visual accessibility in VR by providing (a) high-level natural language scene interpretation powered by a VLM, and (b) detailed low-level object localization tools based on speech, audio, and haptic cues.
The system was evaluated in a user study with 12 BLV participants, who were asked to complete three tasks related to scene understanding, object localization, and object interaction with and without \textsc{EnVisionVR} in a VR scene. 
Participants achieved a significantly higher success rate when locating virtual objects with significantly lower perceived difficulty with \textsc{EnVisionVR}.

This research makes three main contributions.
First, the formative study adds to the existing literature on accessibility barriers for BLV users by emphasizing the lack of functions for scene description and interaction support as a key concern.
Second, to the best of our knowledge, \textsc{EnVisionVR} is the first proof-of-concept system to incorporate detailed VLM-based scene descriptions for real-time visual accessibility in VR, through spatial audio, voice instructions, and speech-based function activation methods.
Third, we offer a set of design implications derived from the system's development process and evaluation to inform visual accessibility design in VR more extensively.




\section{Related Work}

\subsection{Visual Accessibility Design in VR}

In a study conducted by Naikar et al.~\cite{naikar2024accessibility}, 39 out of 106 inspected free VR experiences (36.8\%) lacked accessibility features. Furthermore, users may encounter multiple accessibility barriers in the same context~\cite{creed2024inclusive}. Extensive work has focused on advancing visual accessibility 
in VR to provide a more inclusive experience~\cite{dudley2023inclusive}. Mostly, this has been approached through augmenting visual information \cite{zhao2019seeingvr, masnadi2020vriassist, teofilo2018evaluating} or translating it into audio or haptic feedback \cite{canetroller, kim2020vivr, zhao2019seeingvr, ji2022vrbubble}. 

Outstanding work in the area of augmenting visual information includes the development of tools for magnification, contrast adjustment, color correction, text and display size adjustment, among others. Gear VRF Accessibility~\cite{teofilo2018evaluating}, for instance, provided a framework for developers to adapt zoom, invert colors, and add captions in a VR environment. VRiAssist~\cite{masnadi2020vriassist} supported the user by offering visual assistance based on eye tracking, providing tools like magnification, distortion, colour and brightness correction. SeeingVR~\cite{zhao2019seeingvr} involved a larger set of visual augmentation tools that proved effective for task completion in VR (such as menu navigation, visual search, and target shooting).
Consistent with these approaches, Ciccone et al.~\cite{ciccone2023next} recommended implementing contrast adjustment controls, color correction controls, and font and display size adjustments to increase information visibility when designing for visual accessibility.


Research focused on converting visual information into other forms has also resulted in a variety of systems supporting visual accessibility in VR. For instance, both Canetroller~\cite{canetroller} and VIVR~\cite{kim2020vivr} simulated the use of a white cane in the virtual world. This included providing 3D spatial audio feedback, physical resistance, and vibrotactile feedback to simulate cane--virtual object interaction. The aforementioned SeeingVR~\cite{zhao2019seeingvr} also included text-to-speech and object recognition from visual information to speech. In a more specialized context, Dang et al.~\cite{dang2023opportunities} outlined a multimodal-multisensor VR system with spatial audio, audio descriptions, audio feedback, and vibrotactile feedback to enhance the experience of BLV participants in immersive musical performances. 
Finally, VRBubble~\cite{ji2022vrbubble} enhanced BLV users' peripheral awareness to facilitate social VR accessibility through audio alternatives such as earcons, verbal notifications, and real-world sound.



Among both approaches, augmenting visual information cannot support users who are blind or with very limited visual perception. Thus, the work in this paper focuses on integrating the relatively underexplored methods of converting visual information into speech, audio cues, and haptics. We investigate how VLMs could be incorporated to provide vivid scene descriptions. By combining these multiple modalities, we aim to provide users with a high-level understanding of their surroundings, as well as a detailed understanding of object-level information to support interaction.


\subsection{Screen Readers and Web Accessibility}

Screen readers are a well-established accessibility tool for BLV users; 
their design concepts can provide valuable insights for the design of visual accessibility in immersive environments.
NVDA, JAWS, and VoiceOver are three of the most commonly used screen readers for desktops and laptops~\cite{WebAim-10}. While these different screen readers have distinct characteristics, they share key design principles which underpin their effectiveness. First, popular screen readers prioritize keyboard navigation. 
Keyboard navigation allows users to navigate digital content without the need for a mouse, which is critical for people with vision impairment~\cite{accessibleWebDev2021}. 
Second, screen readers focus on the semantic structure to facilitate smooth navigation and ensure information accuracy. On this topic, a series of works~\cite{zong2022rich, di2004usable, williams2019find} have specifically focused on how to improve the usability of screen readers by correctly and efficiently conveying semantic details. 
Third, screen readers also provide alternative text for images, which is a crucial step to help convey non-textual content~\cite{alttext2017, altText2018}. Fourth, screen readers use headings and landmarks to assist website navigation and hierarchy~\cite{southwell2013evaluation}. Finally, screen readers also assist user input, such as filling in and submitting forms and documents online, an important part of web interaction~\cite{borodin2010more}.


\textsc{EnVisionVR} takes inspiration and expands on the design principles and concepts of screen readers. 
Based on the above, we arrive at an interactive design that uses speech commands as a parallel to keyboard navigation, while constructing high-level scene information and detailed object-level information for BLV users as a parallel to the semantic structure processed by screen readers. Furthermore, 
VLMs provide a highly efficient way to produce scene descriptions, 
a parallel to explicit alternative text.

\subsection{Powering Visual Accessibility with Artificial Intelligence}

The emergence of powerful VLMs has enabled the automated generation of high-quality descriptions of visual information.  
Current VLMs~\cite{salin2022vision,pmlr-v139-cho21a,luo2022vc,bongini2023gpt,zhang2023gpt4mia} are capable of jointly processing images and text data for image captioning, visual question answering, and medical image analysis. 
These models are now being deployed in a range of use cases to power visual accessibility features. 
For example, De La Torre et al.~\cite{de2024llmr} demonstrated potential applications of their Large Language Model (LLM)-based tool for 3D scene editing in visual accessibility. Jiang et al.~\cite{jiang2023beyond} highlighted the potential of advanced AI models to enhance the quantity and quality of audio descriptions.
Microsoft developed SeeingAI~\cite{microsoft-seeing-ai} to narrate the physical world for BLV users.
Similarly, Be My Eyes launched Be My AI~\cite{BeMyAI}, an AI assistant powered by GPT-4, which provides detailed descriptions of photos taken and uploaded by BLV users, and a braille display for deaf-blind users. 
Specific use cases for scene description in real-life scenarios have been identified through a diary study~\cite{gonzalez2024investigating}, which highlights the effectiveness of generative models for visual accessibility design.

The increasing attention to applying VLMs to interactions in 3D content and accessibility design illustrates the strong capability of such models. While existing work has demonstrated how state-of-the-art models could be applied 
in accessibility design for 2D images or videos, there has been limited work studying how these models could be applied in accessibility design for VR immersive environments. 

\section{Formative Study}\label{sec:formative}

In the formative usability study, which involved nine BLV participants, we sought to understand the accessibility barriers encountered 
in consumer-based VR and AR technology. Through this process, we studied the adaptations implemented when facing such barriers, namely the way people with specific access needs modified their behaviour or received assistance from another non-disabled person to fully or partially overcome these issues. 

\subsection{Method}

Our study protocol evaluated the usability of 
representative consumer-level, single-user VR experiences to identify the types of barriers encountered. The majority of the experiences represented content currently available in the market, while others were included to cover the full range of usability demands in VR, such as vision, hearing, touch and physical movement, and interaction modes, such as controller-based and hand-tracking. The study tasks and experiences became progressively more complex as the study progressed. 
See the Online Appendix for the complete set of experiences, tasks and sub-tasks.

The Meta Quest 2 was used. Tasks and experiences were designed following the typical user journey; beginning with wearing and fitting the VR hardware (\textit{VRH:~Headset}, \textit{VRC:~Controllers}), 
followed by navigating the Meta Quest home menu to configure existing accessibility features (\textit{VR1:~Menu}), 
and completing each of the selected experiences: 
\textit{VR2:~“As it is” 360° video}\footnote{Produced by \href{https://www.youtube.com/watch?v=BE-irHmbQOY}{\underline{360 Labs}}} (immersive video documentary),
\textit{VR3:~Job Simulator}\footnote{Produced by \href{https://jobsimulatorgame.com/}{\underline{Owlchemy Labs}}} (videogame simulating a cooking scenario, using virtual hands to manipulate objects while following cooking instructions), 
\textit{VR4:~Moss}\footnote{Produced by \href{https://www.polyarcgames.com/games/moss}{\underline{Polyarc}}} (storyline-based videogame where the user becomes a secondary character that interacts with objects and controls other characters), 
and \textit{VR5:~Elixir}\footnote{Produced by \href{https://www.magnopus.com/projects/elixir}{\underline{Magnopus}}} (hand-tracking-based videogame where the user manipulates virtual objects with their real hands). 
Sub-tasks 
were basic commands revolving around specific steps required to progress through each experience and explore available features and interactables. 

In total, participants completed 34 
sub-tasks spread across two VR hardware tasks and five VR experiences  
(e.g., \textit{`Adjust the focal distance of the headset'}, \textit{`Spot different visitors in the scene, from those close by to those at a distance'}). Participants were asked to perform each sub-task while thinking aloud. A researcher scored task success on a 0--3 scale (0 = unable to start or finish the task, even with adaptations, 1 = able to start but unable to finish the task, even with adaptations, 2 = successful completion of the task with adaptations, and 3 = successful completion of the task without adaptations). The concept of \textit{adaptation} arose after a pilot study that showed most sub-tasks were not achievable for multiple people with access needs. Thus, we resolved to study adaptations as either \textit{self-initiated} unconventional behaviour (e.g., holding a VR controller with two hands for pointing accuracy) 
or \textit{assistance} from another person through a Wizard of Oz approach (e.g., imitating a non-existent screen reader feature). 

The study was approved by the Ethics Committee of the College of Engineering, Design and Physical Sciences, Brunel University of London.
The study session 
lasted approximately 120 minutes per participant. 
The sessions were facilitated by a researcher experienced in providing BLV accessibility support; they were in charge of observing, scoring sub-tasks, and assisting the participants. A technician was in charge of onboarding and looking after the technical elements. 

\subsubsection{BLV Participants}

Nine participants (3 female, 6 male) who self-reported as blind or with low vision were recruited through an inclusive research user panel (managed by Open Inclusion \cite{OpenInclusion}). All participants provided informed consent. 
Their ages ranged from 27 to 68 (\textit{M} = 43.11 years, \textit{SD} = 13.82) and their previous experience with VR technology ranged from novice (1) to competent (3). For these participants, sight was classed as the access need that impacted their lives most extensively. 
These details are summarized in Table~\ref{fig:M-heatmap}. 
To distinguish from participants in the study reported in \Cref{sec:study-design}, participants in the formative study are labelled as PF1 to PF9.

\begin{table*}[!t]
    \centering
    \caption
    [Table presenting the participant demographics and task success scores for the formative study with BLV users. The first column presents the Participants code, from PF1 to PF9. The second column presents the corresponding participant's age. The third column presents the participants' gender. The fourth column presents a description of each participant's vision impairment. The fifth column indicates the level of VR experience before the study (1 representing Novice and 5 representing Expert). The demographics are as follows: PF1, age 56, woman, blind, VR expertise level 2; PF2, age 38, woman, lacks most central vision and right peripheral vision, nystagmus, VR expertise level 3; PF3, age 41, man, blind, VR expertise level: 2; PF4, age 31, woman, light sensitivity due to Autism, VR expertise level: 3; PF5, age 56, man, blind in left eye, only central vision in right eye, VR expertise level: 1; PF6, age 68, man, double vision due to Multiple Sclerosis, VR expertise level: 1; PF7, age 27, man, Irlen Syndrome, VR expertise level: 1; PF8, age 32, man, blind, VR expertise level: 2; and PF9, age 39, man, anoridia with sensitivity to light and glare, nystagmus, VR expertise level: 1. The final eight columns provide the mean scores (\textit{M}) and standard deviation (\textit{S}) for the tasks VRH: Headset and VRC: Controllers, and for the experiences VR1: Menu, VR2: "As it is" 360-degrees video, VR3: Job simulator, VR4: Moss and VR5: Elixir. At the right of the table, a colour gradient is included: from dark red for 0, to light green for 3.00. From this heatmap, is noted that PF1 presented consistently lower scores, followed by PF8. And that the VR2 ("As it is" 360-degrees video) and VR5 (Elixir) were the experiences where participants scored lower.]
    {Participant demographics and task success scores for the formative study with BLV users. VR Expertise indicates VR experience level (1 representing Novice and 5 representing Expert). Mean scores ({M}) and standard deviation ({S}) are provided for each VR task or experience.}
    \label{fig:M-heatmap}
    \includegraphics[width=\linewidth]{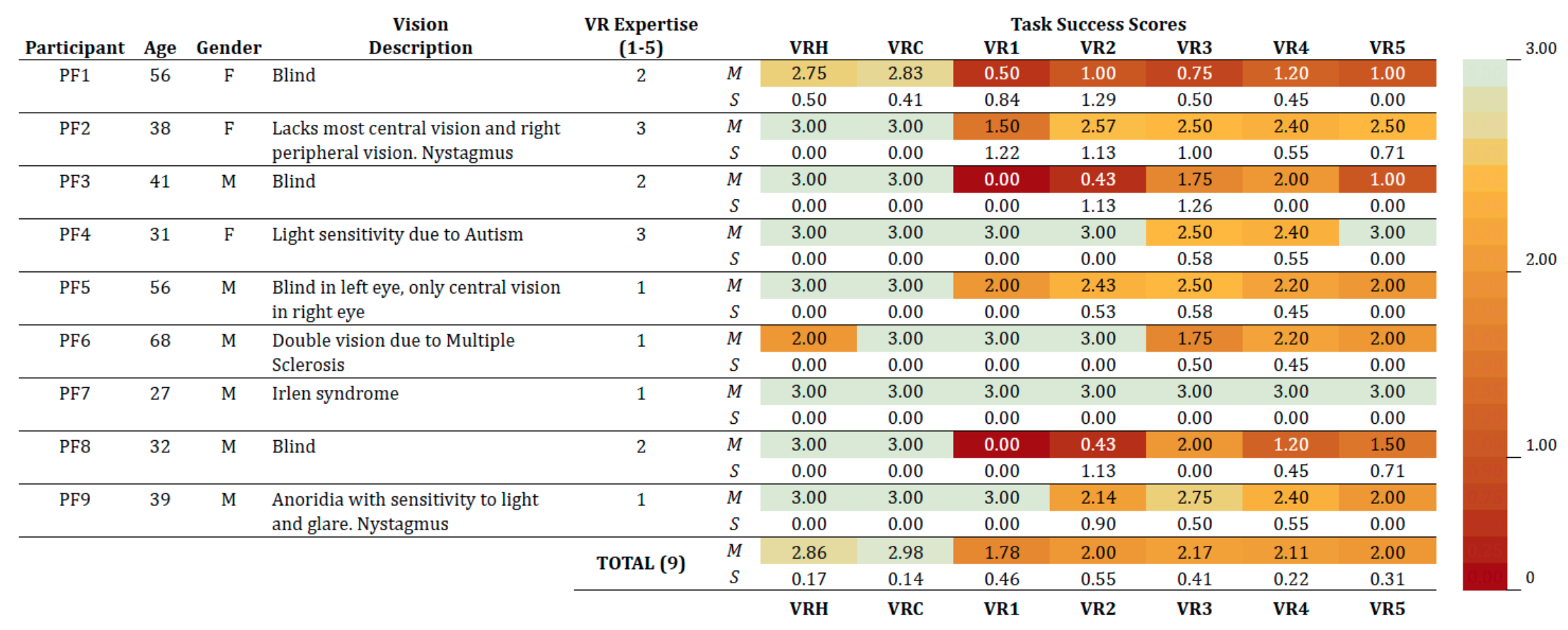}
    \centering
\end{table*}

\subsection{Results}


\subsubsection{Task Success}

This score indicates the level of success in completing a sub-task. Each participant was presented with 34 sub-tasks in total (\textit{VRH} = 4, \textit{VRC} = 6, \textit{VR1} = 6, \textit{VR2} = 7, \textit{VR3} = 4, \textit{VR4} = 5, \textit{VR5} = 2). 
305 individual scores were produced across the nine participants over the seven VR tasks/experiences; one sub-task was not performed due to participant request (PF8, \textit{VRH}). 
Mean scores 
are summarized in Table~\ref{fig:M-heatmap}.
The 
home menu (\textit{VR1}) presented the lowest total mean score 
while the highest was achieved in Job Simulator (\textit{VR3}). 
Low-vision participants could fully or partially complete \textit{VR1} and \textit{VR2} sub-tasks, blind participants were unable to start most of these sub-tasks. This trend partially evened out on the next experiences (\textit{VR3}, \textit{VR4} and \textit{VR5}), with all participants presenting mixed score levels, although blind participants were still positioned in the lower scale. 



\subsubsection{VR Accessibility Barriers}

Whenever a participant scored 2 or below (Task Success) in a sub-task, a usability friction instance was logged. 
A total of 157 instances of usability friction were encountered. 
The most common barrier 
revolved around the lack of screen reader, with a 25.48\% frequency (n~=~40 out of 157 total instances). 
This barrier was closely followed by no zoom or magnification options and issues operating the VR controllers, 
each with a 17.20\% frequency (n~=~27). 
Participants were generally able to adapt to these barriers through external support (n = 142, 90.45\%), such as help provided by the facilitator, and on a few occasions through self-initiated actions (n = 9, 5.74\%). 
Common adaptations included requesting further instructions to interact within the specific VR environments, with a 24.84\% frequency (n~=~39), closely followed by `mimicking' missing assistive features such as an \textit{ad-hoc} screen reader (n~=~35, 22.29\%) or \textit{ad-hoc} audio descriptions (n~=~31, 19.75\%). 

\subsubsection{Adaptations for Blind Participants}

Blind participants had difficulty with experiences that only provided single-modality outputs. 
This issue was particularly notable when information was only communicated by visual means, but participants also experienced some difficulty interpreting information presented in a single modality using audio or haptics.
Often, audio descriptions of the play space, interactable objects and the VR pointer location were necessary to help them orient themselves. 
This was most common at the beginning of the experience, and when haptic and audio signals were not enough to convey the type of the object (PF1, PF3, PF8).

Audio tutorials were also helpful when friction occurred. On most occasions, tutorials guided participants on controller use, for example to explain how controllers were mapped to interactions in a specific scene, or how to manipulate interactables or control characters 
(PF1, PF3, PF8).
In this regard, PF3 highlighted the need for a directional cueing system that could, for instance, guide them to move their controllers closer to the menu.

Audio feedback in combination with haptic feedback was another requirement identified throughout the study. When they were provided conjointly (e.g., \textit{VR3} used haptics and audio to simulate the opening of a virtual door), PF1 and PF3 could more easily perceive what was going on in the scene. When such signals were poor or did not exist, it became more difficult for participants to orient themselves 
(PF1, PF3 PF8).


\subsubsection{Adaptations for Low-Vision Participants}

Low-vision participants such as PF5 completed more sub-tasks than the blind participants but required longer periods to familiarize themselves with the virtual environments.
Explicit audio descriptions and tutorials were important to clarify what participants were partially seeing in a scene. PF9, for instance, benefited from audio description in \textit{VR2.} 
Detailed and repeated instructions were helpful for PF2 in \textit{VR5}. 


Multimodal feedback was important in some instances. For PF2, multiple signals (i.e. peripheral vision, haptics and sound) helped them to touch the interactable objects in \textit{VR3}. 
But they struggled in \textit{VR2} because there was no haptic feedback confirming button interaction. 
Similarly in \textit{VR4}, several participants encountered difficulty because there was no clear signal when an object was interactable and the colour contrast was low (PF2, PF5, PF9).

\begin{figure*}[ht!]
  \includegraphics[width=0.33\textwidth]{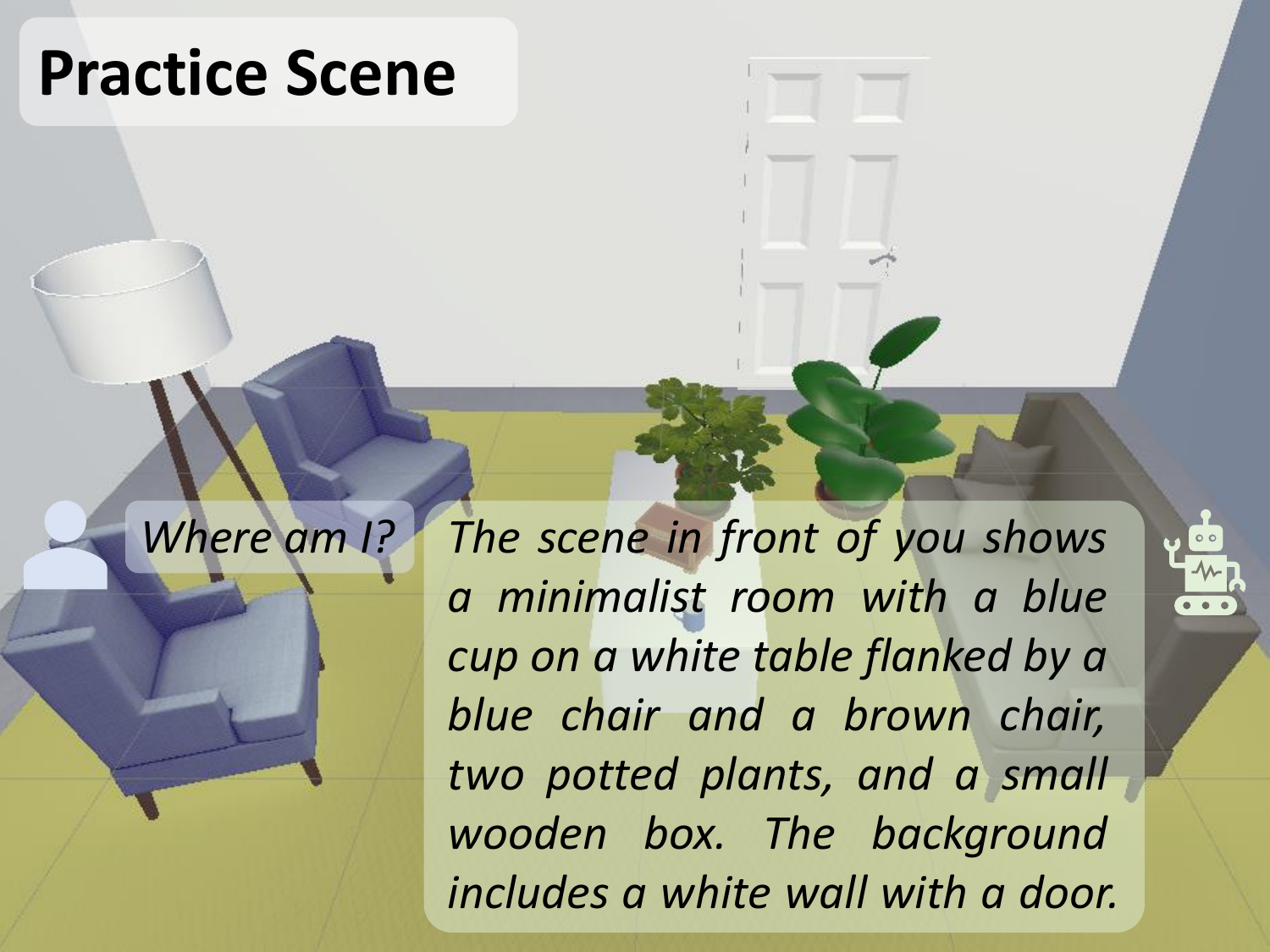}
  \includegraphics[width=0.33\textwidth]{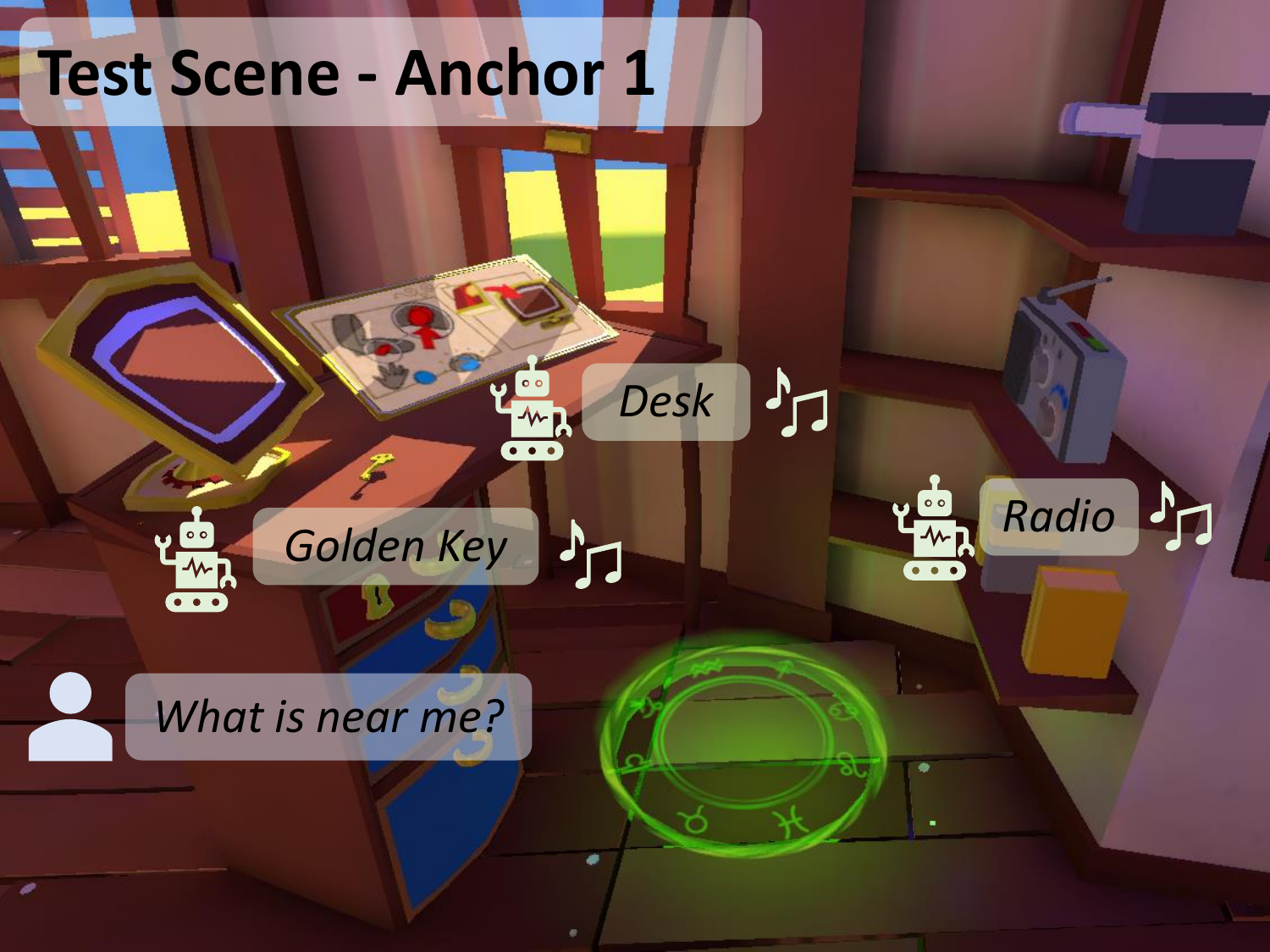}
  \includegraphics[width=0.33\textwidth]{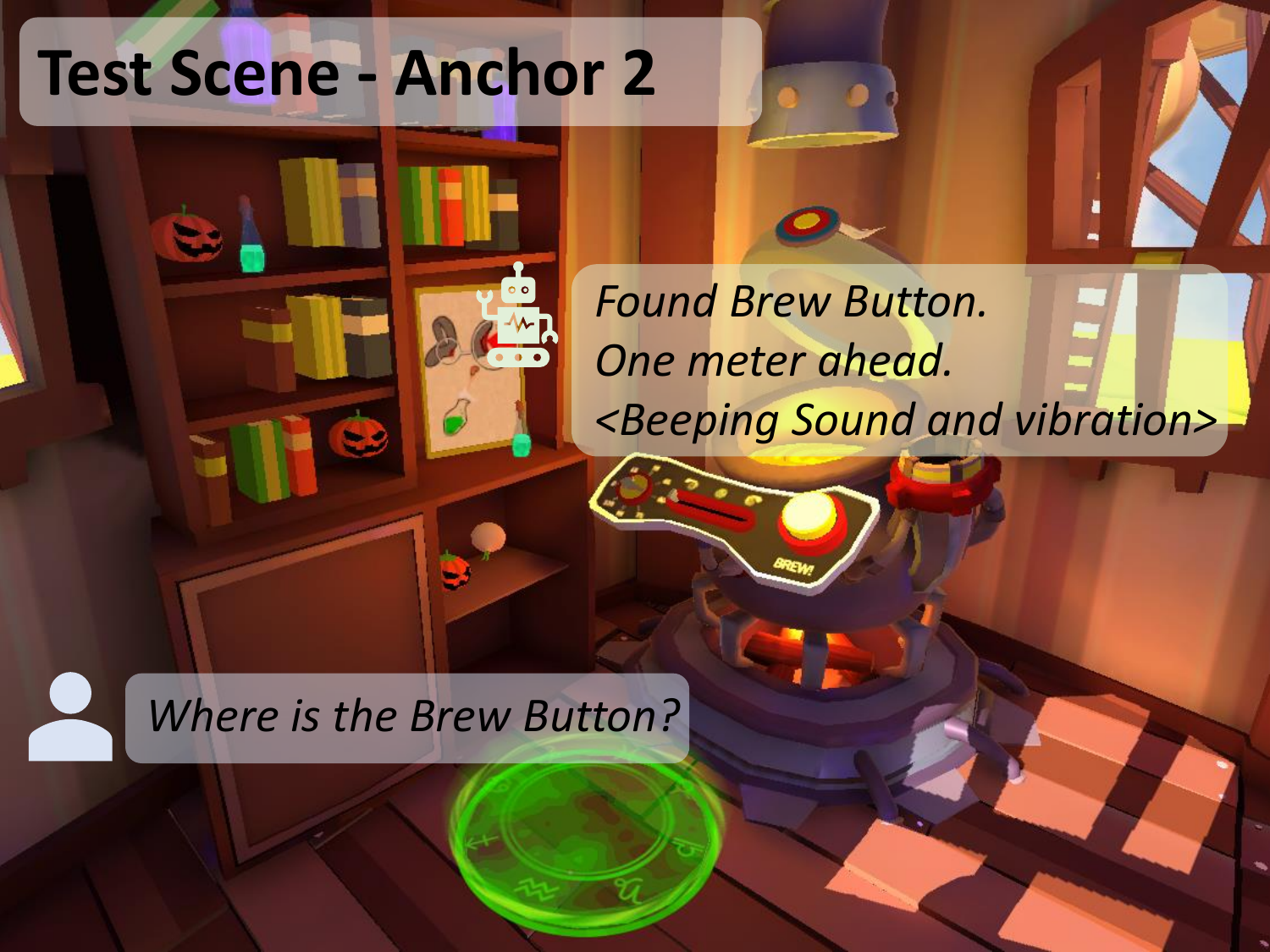}
  \caption{Examples of functions supported by \textsc{EnVisionVR} to enhance the accessibility of VR experiences for BLV users. Left: The user can ask ``Where am I?'' and the \textsc{EnVisionVR} system reads out a detailed description of the user's current field of view. Middle: The user can ask ``What is near me?'' and the system reads out the names of the three main objects near the user with a spatial tone to indicate the object's location. Right: The user can ask ``Where is the Brew Button?'' and the system uses a beeping sound and directional instructions to communicate the distance to the Brew Button. When the user reaches the Brew Button, the controller vibrates to inform the user.}
  \label{fig:teaser}
\end{figure*}

\subsection{Summary}

Results from the formative study revealed that BLV users face various accessibility barriers in VR. The lack of integrated screen reader functionality and audio descriptions were identified as major barriers preventing them from completing sub-tasks without assistance. While low-vision participants required longer periods to familiarize themselves with the environments, or concrete audio descriptions to clarify the scenes, blind participants were unable to carry out specific sub-tasks in \textit{VR4} and \textit{VR5} because there was no appropriate multimodal feedback to, for instance, understand the location of objects. In contrast, \textit{VR3} offered helpful audio cues when reaching interactable objects. 

The provision of accessibility features was irregular across the experiences, consistent with the findings of Naikar et al. \cite{naikar2024accessibility}. BLV participants continually struggled to complete sub-tasks related to visual capability demands, revealing that where visual accessibility features existed (e.g., colour contrast adjustment, audio levels) these were insufficient. 
It was also observed that isolated screen reader or audio description implementations are unlikely to accommodate BLV users’ requirements. In contrast, a more complex system dedicated to providing high-level descriptions of the virtual environments coupled with detailed descriptions of interactable objects could serve as an effective guide both for blind and low-vision users facing difficulty with navigating VR experiences. 

\section{Design of \textsc{EnVisionVR}}

The formative study suggests that BLV users require a scene interpretation functionality which builds upon the principles of a screen reader to incorporate spatial elements to assist users to navigate and interact within the 3D space. 
While prior work, such as \textsc{SeeingVR} \cite{zhao2019seeingvr}, has focused on features that support visual perception (e.g. zooming, contour highlighting), we pursue an alternative strategy and seek to replicate and evaluate the familiar experience of using a screen reader and listening to audio descriptions to provide visual accessibility for 3D content.

In the development of \textsc{EnVisionVR}, we sought to adhere to the following key design objectives.
First, we recognize that the inclusion of visual accessibility features should not necessitate the onerous manual reconfiguration of existing VR experiences.
This requirement acknowledges the current meager state of visual accessibility among existing VR applications as noted in the formative study and by Naikar et al. \cite{naikar2024accessibility}.
For a visual accessibility solution to be widely adopted by developers it should therefore not add substantially to the effort or cost of development and release.
Second, the accessibility features should be minimally disruptive to the user's enjoyment of the primary VR experience.
This objective poses certain constraints such as avoiding the remapping of controller buttons that might be used in the VR experience, or inadvertently introducing new access barriers, e.g., enforcing the use of two controllers or difficult-to-press buttons. 
Third, we seek to not only support BLV users in perceiving the virtual environment but also to interact with virtual objects in the environment.
This principle implies the need to provide information to the user at different levels of granularity, extending from high-level scene descriptions to detailed object-level information. 

In response to these design objectives, we developed \textsc{EnVisionVR}, a generalized visual accessibility framework which can be deployed into a VR application with minimal integration effect.
This framework consists of: (i) the Scene Description Function; (ii) the Main Objects Indication Function; and (iii) the Object Localization Function.
To minimize the need for remapping controller buttons, these functions can be activated by three simple speech commands, namely ``Where am I?'', ``What is near me?'', and ``Where is the $<$object name$>$?''. 
For the implementation evaluated in this paper, the user must first press Button A to issue a voice command but this could in theory be changed to a `wake' word or remapped to any other button.
The use of these three functions is illustrated in \Cref{fig:teaser} and their implementation is described in more detail in the remainder of this section.


\subsection{Scene Description -- \textit{Where am I?}}

\begin{figure}[ht!]
    \centering
    \includegraphics[width=\linewidth]{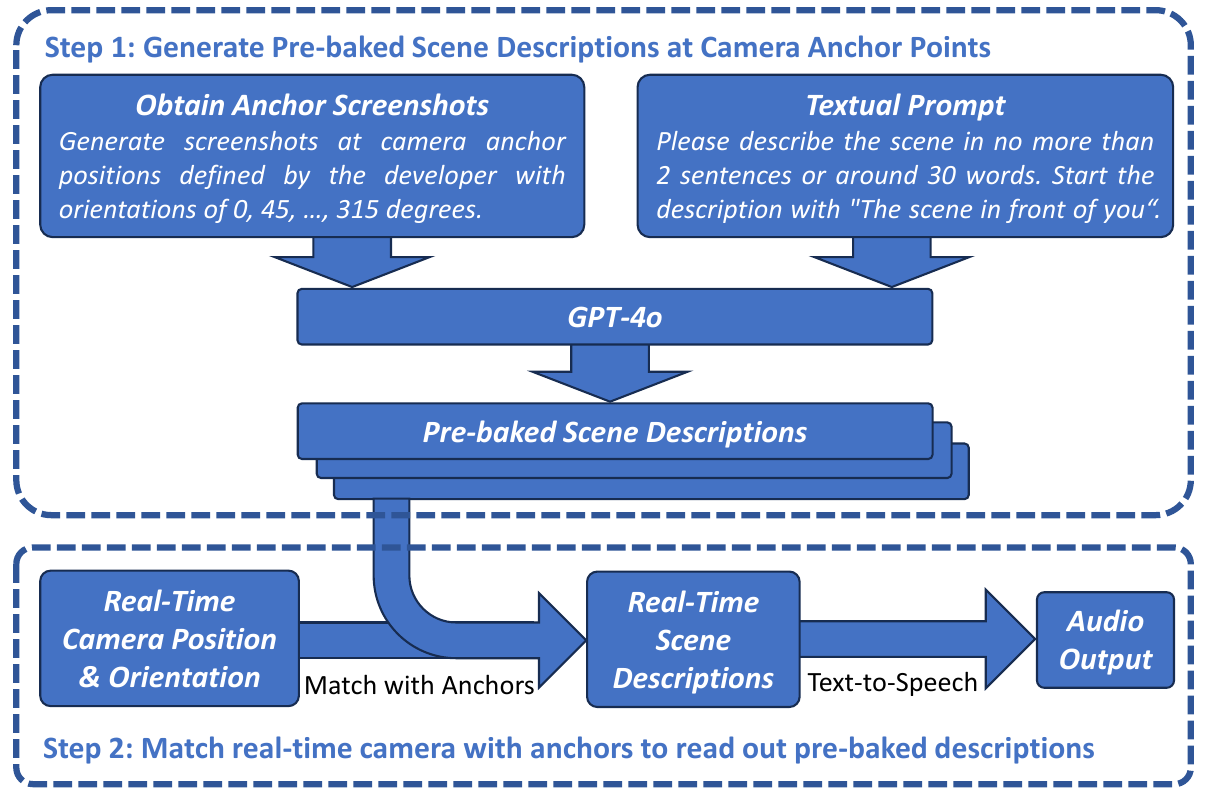}
    \caption{Overview of the Scene Description Function. Scene description is provided in two steps. In Step 1, camera anchor positions are determined by the developer or automatically by the system. Screenshots of the field of view of these anchor points with orientations of 0, 45, ..., 315 degrees along the horizontal plane together with a textual prompt are fed into GPT-4o to generate pre-baked scene descriptions. In Step 2 during runtime, we match the current camera position and orientation with the closest-matching anchor position and orientation to read out the pre-baked descriptions via the Microsoft text-to-speech (TTS) service.}
    \label{fig:scene-description-methodology}
\end{figure}

Issuing the \textbf{``Where am I?''} voice command triggers the \textbf{Scene Description Function}, which describes the user's field of view in a few sentences. 
An overview of the implementation of the Scene Description Function before and during runtime is provided in \Cref{fig:scene-description-methodology}. Details of each step in the implementation are provided in the Online Appendix. 

\begin{figure}[ht!]
    \centering
    \includegraphics[width=\linewidth]{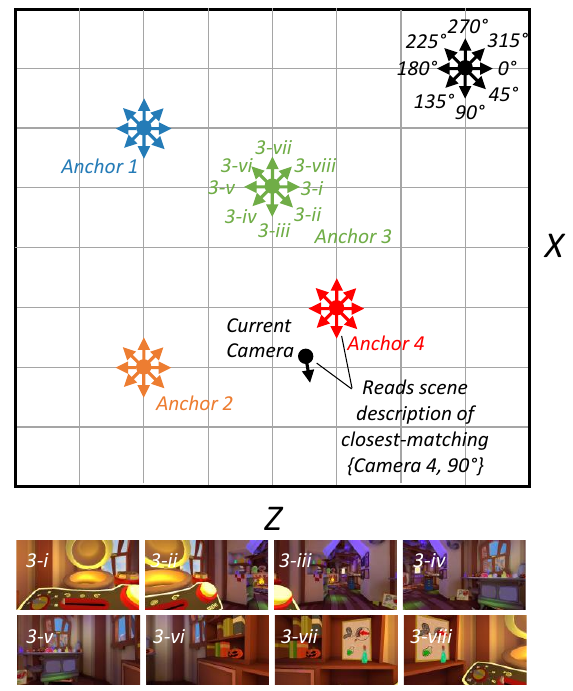}
    \caption{Top-down view of camera anchor positions in a VR escape room and the user field of view in eight directions for each anchor point (left). At each field of view, a screenshot is taken to generate the pre-baked scene description. Example field-of-view screenshots taken at Anchor 3 are provided in the bottom images.}
    \label{fig:anchor}
\end{figure}

Before runtime, camera anchor points\footnote{We define \textit{anchor points} as a list of $(x, y, z)$ coordinates which define the position, but not the orientation, for the user camera to be placed in the scene, such that the user camera placed at all anchor points, with eight different orientations each, capture user field of views with all of the important objects in the scene.} are determined manually by the developer.
As shown in \Cref{fig:anchor}, upon specifying the camera anchor points, a script is executed to automatically capture eight screenshots of the user field of view at each anchor point with orientations of 0, 45, ..., 315 degrees. These screenshots are then sent to a vision language model (VLM) together with a textual prompt, and a short scene description is obtained for each camera anchor position and preset orientation. For example, if four camera anchor positions are determined, $4\times8=32$ scene descriptions are generated. 
We used GPT-4o as the VLM and used the textual prompt ``Please describe the scene in no more than 2 sentences or around 30 words. Start the description with `The scene in front of you''' to generate the scene descriptions. The scene descriptions are stored locally in a CSV file.
As the scene descriptions are generated before users execute the application, we refer to these descriptions as \textit{`pre-baked'} \footnote{Currently, scene descriptions are generated prior to users executing the application (i.e. pre-baked). If the VR scene is modified during runtime the scene descriptions will not be accurate.} .

During runtime, the Scene Description Function processes the current user position and orientation to find the anchor point with the closest matching position and orientation and reads out its `pre-baked' scene description using the Microsoft Azure text-to-speech (TTS) service. As the scene descriptions have been generated by the VLM before runtime, this mapping process allows scene descriptions to be read out to the user with very low latency, usually within tens of milliseconds ($M=19.8, SD=19.5$) based on data from our user study.

\subsection{Main Objects Indication -- \textit{What is near me?}}

Issuing the \textbf{``What is near me?''} voice command triggers the \textbf{Main Objects Indication Function}, which announces the names of three key objects which are near the user, each followed by a short spatial tone to indicate the object's location relative to the user headset in the scene.
Exactly which objects are read out is determined by a `runtime importance value'. This value is proportional to a preset importance value,
and inversely related to the distance between the object and the user camera.
Here, the preset importance value for all objects can be determined automatically by \textsc{EnVisionVR} based on the presence of rendering components,
or specified manually by the developer.
If an object has been previously announced, its runtime importance value is reduced to allow other objects to be announced in subsequent activations of the function. Details on how the importance values are determined can be found in the Online Appendix.

\subsection{Object Localization -- \textit{Where is the <object name>?}}

Issuing the \textbf{``Where is the $<$object name$>$?''} voice command triggers the \textbf{Object Localization Function}, which starts a beeping sound with the beeping frequency inversely proportional to the distance between the right controller and the object. Directional and distance information (e.g., `1 meter ahead') is also provided at regular time intervals to guide the user. When the controller is close enough to the object to interact with it, the controller vibrates.
If the object is interactable and can be held, the system will also announce ``holding $<$object name$>$'' when it is picked up. More implementation details are provided in the Online Appendix.

\section{Evaluation Study}\label{sec:study-design}

To evaluate the potential benefits of \textsc{EnVisionVR}, we conducted a user study with 12 BLV participants.
Participants completed three types of prescribed tasks in VR both with \textsc{EnVisionVR} and without.
The without condition represented the default experience available to BLV users with no dedicated visual accessibility features.
The study was approved by the research ethics committee in the Department of Engineering at the University of Cambridge.

\subsection{Method}

A within-subjects design was adopted to evaluate the performance of \textsc{EnVisionVR} (abbreviated in \Cref{sec:results} as \textsc{EVR}) and the no accessibility features condition (abbreviated in \Cref{sec:results} as \textsc{NVR}).
The order of conditions was counterbalanced. 
For each condition, participants were first familiarized with the available functions in a practice scene (see \Cref{fig:teaser} left image).
Participants were encouraged to use all available functions and the experimenter gave examples of the types of tasks they would be asked to complete. 
After completing familiarization in the practice scene, participants were transported to the test scene. 

The test scene, a VR Escape Room~\cite{vr-escape-room}, was chosen for several key reasons.
First, it is a tutorial scene made freely available by Unity and so provides an example of the type of existing VR experience that may require retrofitting of accessibility features.
Second, it contains rich objects and scene elements.
Third, it supports different interactions with virtual objects (such as grabbing objects and pressing buttons). 
We define two study anchor points in this scene (see \Cref{fig:teaser} middle and right image) and the combination of condition and anchor point is balanced across participants. 
At the given study anchor point, participants then completed the tasks described in the following subsection.

\subsection{Tasks}\label{sec:tasks}

The degree to which \textsc{EnVisionVR} supports BLV users in perceiving and interacting with the virtual environment is evaluated across the three tasks summarized below.
A complete list of questions and tasks for the two anchor points in the test scene is provided in the Online Appendix.

\vspace{0.3em}

\begin{enumerate}
    \item \textbf{Scene Understanding Task:} Participants are asked to rate a statement to evaluate their understanding of the scene from 1 (very unlikely to be true) to 5 (very likely to be true). For example, at Anchor 1 (see \Cref{fig:teaser} middle image), participants are asked to judge whether the statement ``This is a scene of a classroom with a desk and a chair'' is likely to be true or not. 
    \item \textbf{Object Localization Task:} Participants are asked to turn to face a specified object in the scene. For example, they are asked to turn to face the radio at Anchor 1. The task completion status was recorded as a yes/no binary value. 
    \item \textbf{Object Interaction Task:} Finally, participants are asked to interact with an object in the scene. For example, they are asked to push the ``Brew Button'' at Anchor 2 (see \Cref{fig:teaser} right image). Again, we record their task completion status as a binary value. 
\end{enumerate}

\vspace{0.3em}

For each task, participants also rated the difficulty they encountered in completing the task on a scale from 1 to 5.

\subsection{Participants} 
We recruited a new participant sample with the assistance of Open Inclusion \cite{OpenInclusion}. All participants provided informed consent. The sample consisted of 12 participants, of which three reported being blind and nine reported having low vision. All three blind participants reported regular use of screen readers or other forms of assistive technology. Among the nine participants who reported having low vision, five participants reported regular use of assistive technology, yielding a total of eight participants who regularly use assistive technology. 
\Cref{tab:demographics} provides a summary of the collected demographic information of all 12 participants. To differentiate from the formative study, participants are labeled as P1 to P12.

\subsection{Apparatus}
During the experiment, participants wore a Meta Quest 3 headset and held the right controller.
Participants completed all tasks while remaining seated in a swivel chair.
The headset was connected to a Windows 11 laptop 
in wired `link' mode.

\begin{table*}[ht!]
    \caption{Participant demographics for the evaluative study with BLV users.}
    \label{tab:demographics}
    \renewcommand{\arraystretch}{1.5}
    \resizebox{\textwidth}{!}{
    \begin{tabular}{p{0.1\textwidth}>{\centering}p{0.028\textwidth}>{\centering}p{0.06\textwidth}>{\centering}p{0.09\textwidth}>{\centering}p{0.12\textwidth}>{\centering}p{0.06\textwidth}>{\centering}p{0.06\textwidth}>{\centering}p{0.24\textwidth}>{\centering}p{0.24\textwidth}>{\arraybackslash}p{0.06\textwidth}}
    \hline
    \textbf{Participant} & \textbf{Age} & \textbf{Gender} & \textbf{Education}               & \textbf{VR Experience}                         & \textbf{Vision}     & \textbf{From Birth} & \textbf{Vision Description}                                                                                              & \textbf{Assistive Technology}                                                                                               & \textbf{Regular Use} \\ \hline
    P1              & 58  & Female & Masters                 & Inexperienced                         & Blind      & No         & Sighted in the past but have no usable vision today.                                                            & Voiceover and Jaws as screen readers; Other tech with audio assistance at home.                                    & Yes         \\
    P2              & 21  & Female & A levels                & Inexperienced                         & Low Vision & Yes        & Born with cataracts and glaucoma, able to see a decent amount with glasses.                                     & Uses phone and screen magnifiers to zoom in.                                                                       & No          \\
    P3              & 73  & Male   & GCSE                    & Highly Inexperienced                  & Blind      & No         & Lost sight gradually, totally blind for the past 2 years.                                                       & Uses screen reading software: JAWS, NVDA, Voiceover.                                                               & Yes         \\
    P4              & 50  & Female & Higher National Diploma & Inexperienced                         & Low Vision & No         & Stargardt's which affects central vision.                                                                       & Uses Voice Over and Zoom Text.                                                                                     & Yes         \\
    P5              & 79  & Male   & GCSE                    & Highly Inexperienced*                 & Blind      & No         & Lost sight during a degree course.                                                                              & Siri, Alexa, Be My Eyes, JAWS, Voiceover, and other screen readers.                                                 & Yes         \\
    P6              & 36  & Male   & College                 & Neither inexperienced nor experienced & Low Vision & N/A        & Can see 1 meter ahead, central vision in one eye only.                                                          & Phone has voice over - Apple iPhone. Windows PC, Samsung tablet. Talking TV. Uses Seeing AI - to read bus numbers. & Yes         \\
    P7              & 58  & Male   & Postgraduate degree     & Inexperienced                         & Low Vision & N/A        & No sight in left eye, limited central vision (3/60) in right eye. Has ADHD.                                     & Screen magnification user on the computer                                                                          & No          \\
    P8              & 36  & Female & AS Level                & Highly Inexperienced                  & Low Vision & N/A        & Has light perception and no residual vision. Difficulties in reading if the text is not in the right format.    & Uses screen reader on a daily basis. NVDA on laptop. Talkback on Android. Previously iPhone.                       & Yes         \\
    P9              & 41  & Male   & Bachelor's degree       & Highly Experienced                    & Low Vision & N/A        & Little vision in right eye, can see light and dark and the shape of things.                                     & Text enlarger on mobile and computer, and Dragon Naturally Speaking for speech input and feedback                  & No          \\
    P10             & 45  & Male   & Bachelor's degree       & Highly Inexperienced                  & Low Vision & N/A        & Zero sight in left eye, right eye is a prosthetic, 6/36 vision with changing field. Only sees shape and colour. & Has used lots of tech. Doesn't use an actual screen reader. Has reading glasses and uses audiobooks a lot.        & Yes         \\
    P11             & 54  & Female & Bachelor's degree       & Highly Inexperienced                  & Low Vision & N/A        & Sight in right eye, no sight in left eye. Born with cataracts. Can read some print with glasses.                & Does not use audio on the computer. Does not use specific software. Can enlarge print.                             & No          \\
    P12             & 57  & Female & Entry level 2 English   & Highly Inexperienced                  & Low Vision & N/A        & No central vision and a tiny bit of peripheral vision. Can see light, dark, and some outlines.                  & NVDA, Alexa in the house, Android mobile with Synaptec for screen reader.                                          & Yes         \\ \hline
    \end{tabular}
    }
    \vspace{1ex}
    
    {\raggedright \footnotesize  * P5 was new to the concept of VR. He connected VR with soundscapes and gave himself a VR experience rating of `Neither inexperienced nor experienced'. He also said that he had never used technology of this kind later in the testing session, suggesting that an accurate rating could have been `Highly inexperienced'. \par}
\end{table*}

\section{Results}\label{sec:results}

In this section, we first present the study results for each task outlined in Section~\ref{sec:tasks}.
Later in Subsections~\ref{ssec:behaviors} and \ref{ssec:interview} we report on observations of usage behavior with EnVisionVR, as well as qualitative feedback captured in the post-study interview.


\subsection{Scene Understanding Task}

\begin{figure}[ht!]
    \centering
    \includegraphics[height=6.95cm]{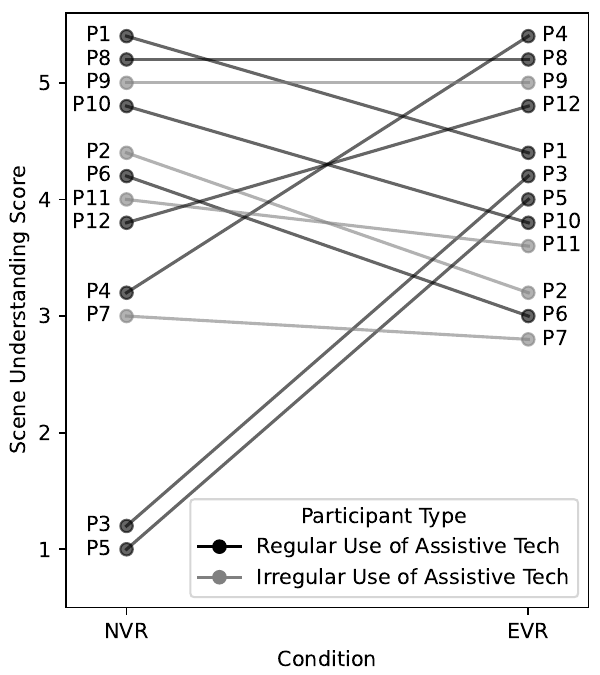}
    \includegraphics[height=6.95cm]{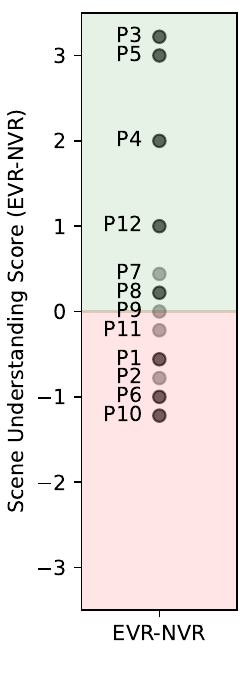}
    \caption[Parallel coordinate plot of the scene understanding task results.]{Scene Understanding Task: Performance of all participants (left) and the difference between scores in the \textsc{EVR} and \textsc{NVR} condition for each participant (right). Participants with blindness and severe visual impairment who regularly use assistive technology are colored in black, while others are colored in grey. Vertical jittering is applied to visualize all points. Participant IDs are labelled beside each scatter point.}
    \label{fig:score-su-parallel}
\end{figure}

\begin{figure}[ht!]
    \hspace{-0.1in}
    \includegraphics[width=\linewidth]{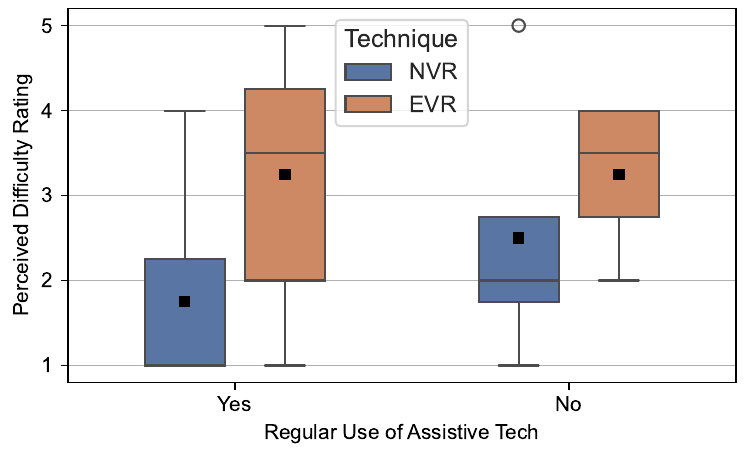}
    \caption[Perceived difficulty of the scene understanding task.]{Scene Understanding Task: Distribution of the perceived difficulty (higher score indicates lower perceived difficulty) for the \textsc{NVR} and \textsc{EVR} conditions for participants who regularly use assistive technology and for those who do not. Black squares indicate the mean value.}
    \label{fig:score-su-difficulty-box}
\end{figure}

In the Scene Understanding Task, participants responded to the given statement on a scale from 1--``very unlikely" to 5--``very likely".
Since at one anchor location, the statement was false, we converted these raw responses such that a higher score indicates a closer match to the correct answer.
\Cref{fig:score-su-parallel} plots the scene understanding score of all participants in the \textsc{NVR} ($M=3.67, SD=1.44$) and \textsc{EVR} ($M=4.08, SD=.793$) conditions.
In \Cref{fig:score-su-parallel} we also make a distinction between whether participants regularly use screen readers or other assistive technology. This roughly groups the full participant group into two subsets based on the degree to which they can directly perceive visual content. 
P3, P4, P5 and P12 who regularly use assistive technology gained a better understanding of the scene with \textsc{EnVisionVR} compared with the condition without any accessibility features. P1, P2, P6, and P10 were able to understand the scene better without \textsc{EnVisionVR}, while P7, P8, P9, and P11 achieved the same level of scene understanding with and without the tool. The decrease in scene understanding performance was due to different reasons such as a lack of attention to long descriptions and failure to capture keywords to support user judgment (P1), or the lack of evidence to convince them to negate the statement which claims the escape room is a classroom (P2, P6). The first-person view descriptions provided only fragments of information about objects around the user, which was insufficient to infer high-level information such as the type and purpose of the scene (P10).

A Friedman test did not indicate a significant difference in scene understanding scores ($\chi^2<.01, p=1.0$) 
between the \textsc{NVR} and \textsc{EVR} conditions.
Friedman tests also did not indicate a significant difference in scene understanding scores between the \textsc{NVR} or \textsc{EVR} conditions for participants who regularly use assistive technology ($\chi^2=.143, p=.705$) or for those who do not ($\chi^2=1.0, p=.317$).

\Cref{fig:score-su-difficulty-box} presents boxplots of the 
perceived difficulty ratings (higher score indicates lower perceived difficulty) of the scene understanding question for the \textsc{NVR} ($M=2.00, SD=1.35$) and \textsc{EVR} condition ($M=3.25, SD=1.29$). 
The difficulty ratings are grouped for participants who regularly use assistive technology (\textsc{NVR}: $M=1.75, SD=1.16$; \textsc{EVR}: $M=3.25, SD=1.49$) and participants who do not (\textsc{NVR}: $M=2.50, SD=1.73$; \textsc{EVR}: $M=3.25, SD=.957$).
A Friedman test did not indicate a significant difference in perceived difficulty ($\chi^2=3.60, p=.058$)
between the \textsc{NVR} and \textsc{EVR} conditions,
or a significant difference in perceived difficulty between the \textsc{NVR} and \textsc{EVR} conditions for participants who regularly use assistive technology ($\chi^2=3.57, p=.059$) or those who do not ($\chi^2=.333, p=.564$).

\subsection{Object Localization Task}

\begin{figure}[t!]
    \centering
    \includegraphics[height=6.95cm]{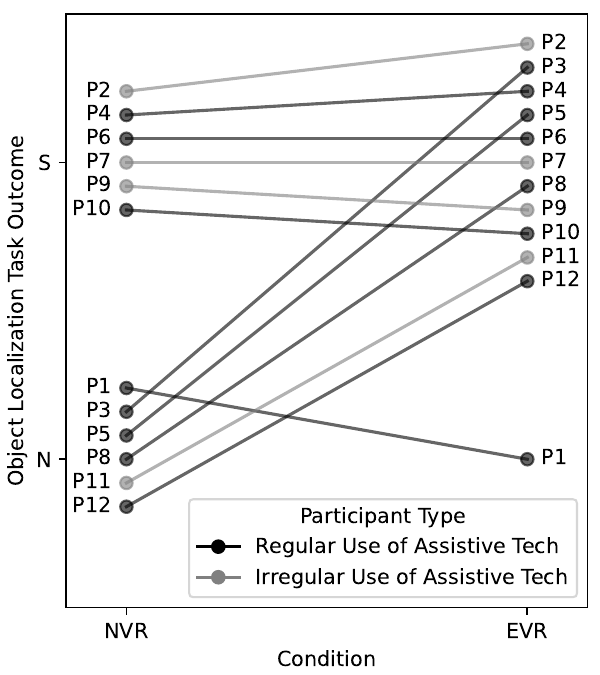}
    \includegraphics[height=6.95cm]{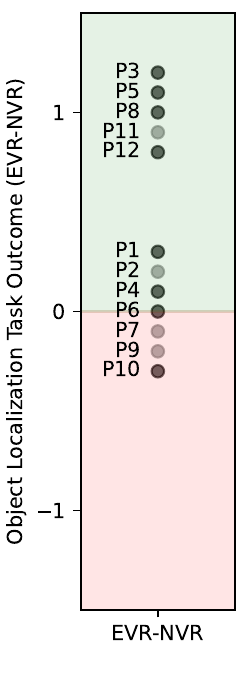}
    \caption[Parallel coordinate plot of the object localization task performance.]{Object Localization Task: Performance of all participants (S: Successful, N: Not successful) (left) and the difference between task outcomes in the \textsc{EVR} and \textsc{NVR} condition for each participant (right). Participants with blindness and severe visual impairment who regularly use assistive technology are colored in black, while others are colored in grey. Vertical jittering is applied to visualize all points. Participant IDs are labelled beside each scatter point.}
    \label{fig:score-ol-parallel}
\end{figure}

\begin{figure}[ht!]
    \centering
    \hspace{-0.1in}
    \includegraphics[width=\linewidth]{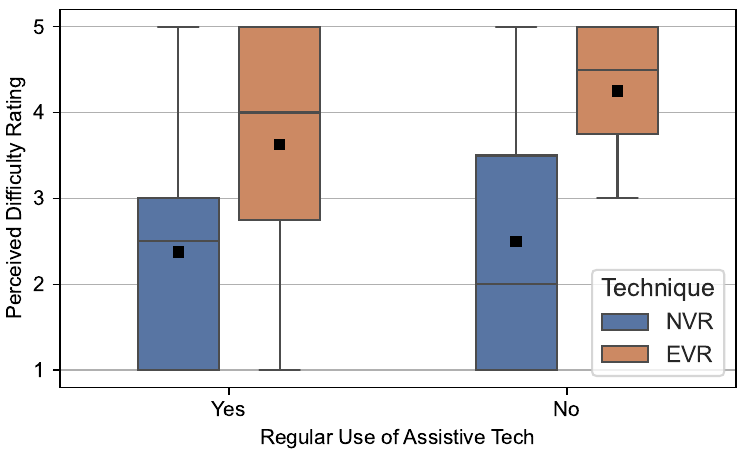}
    \caption[Perceived difficulty of the object localization task.]{Object Localization Task: Distribution of the perceived difficulty (higher score indicates lower perceived difficulty) for the \textsc{NVR} and \textsc{EVR} conditions for participants who regularly use assistive technology and for those who do not. Black squares indicate the mean value.}
    \label{fig:score-ol-difficulty-box}
\end{figure}

\Cref{fig:score-ol-parallel} summarizes the completion status of the object localization task for all participants. 
Six participants were able to complete the object localization task to turn to face a specified object in the NVR condition, while the other six participants were not. In the \textsc{EVR} condition, five of the participants who were unable to complete the task in the \textsc{NVR} condition were able to complete the object localization task.
Only one participant (P1) was still unable to complete the task in the \textsc{EVR} condition, and none of the participants had a worse performance with \textsc{EnVisionVR}. Most participants (3 out of 4) who do not regularly use assistive technology were able to complete the object localization task with or without \textsc{EnVisionVR}, and \textsc{EnVisionVR} was able to help four out of five participants who do regularly use assistive technology, and who could not locate the virtual object in the \textsc{NVR} condition to complete the task. Overall, object localization task completion results show a $91.7\%-50\%=41.7\%$ improvement in task success rate with \textsc{EVR} compared with \textsc{NVR}.
As the object localization task has binary performance data, a McNemar's test was adopted. The test indicated a significant difference ($\chi^2=5.0, p<.05$) between the \textsc{NVR} and \textsc{EVR} task completion status, suggesting that \textsc{EnVisionVR} significantly improved participants' ability to locate virtual objects and significantly reduced their perceived difficulty. A McNemar's test also indicated a significant difference ($\chi^2=4.0, p<.05$) between the \textsc{NVR} and \textsc{EVR} task completion status for participants who regularly use assistive technology, but the difference was not significant ($\chi^2=1.0, p=.317$) for those who do not regularly use assistive technology.

\Cref{fig:score-ol-difficulty-box} presents box plots of
the perceived difficulty of the task for the \textsc{NVR} ($M=2.42, SD=1.51$) and \textsc{EVR} ($M=3.83, SD=1.34$) conditions.
A Friedman's test indicated a significant difference ($\chi^2=4.45, p<.05$) between the \textsc{NVR} and \textsc{EVR} condition for all participants, but this difference was not significant at the subgroup level, i.e. participants who regularly use assistive technology ($\chi^2=3.57, p=.059$) and those who do not ($\chi^2=1.0, p=.317$).

\subsection{Object Interaction Task}

\begin{figure}[ht!]
    \centering
    \includegraphics[height=6.95cm]{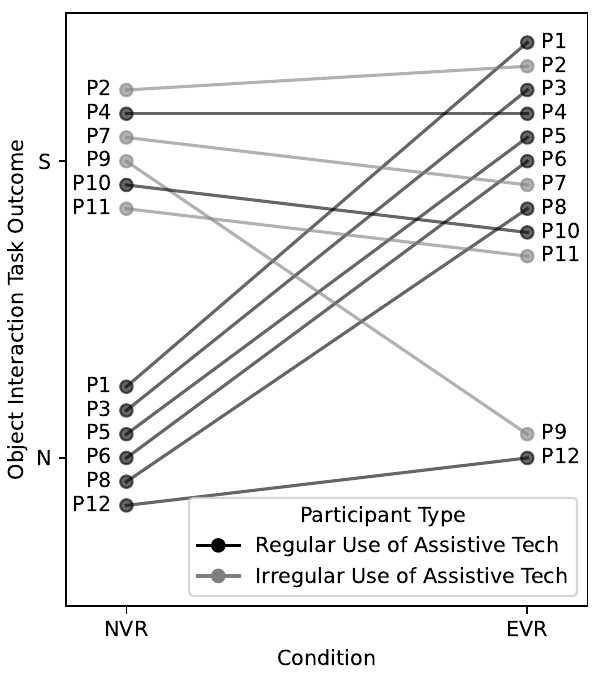}
    \includegraphics[height=6.95cm]{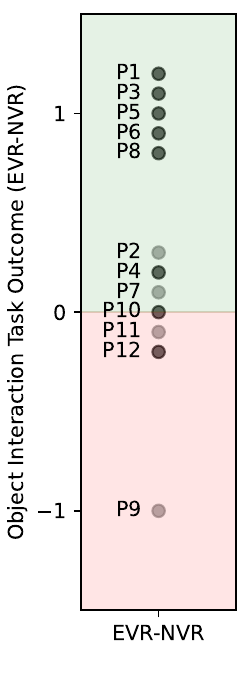}
    \caption[Parallel coordinate plot of the object interaction task performance.]{Object Interaction Task: Performance of all participants (S: Successful, N: Not successful) (left) and the difference between task outcomes in the \textsc{EVR} and \textsc{NVR} condition for each participant (right). Participants with blindness and severe visual impairment who regularly use assistive technology are colored in black, while others are colored in grey. Vertical jittering is applied to visualize all points. Participant IDs are labelled beside each scatter point.}
    \label{fig:score-oi-parallel}
\end{figure}

\Cref{fig:score-oi-parallel} shows the completion status of the object interaction task. Six participants were able to interact with a virtual object (such as picking up a key or pressing a button) under the \textsc{NVR} condition, while the other six were not. Among those who were unable to interact with virtual objects, five participants were able to complete the task with \textsc{EnVisionVR}, while one participant (P12) was still unable to complete the task.
It is worth noting that P12 reported a secondary access need based on her learning disability, and this may have contributed to the difficulty they experienced in completing the task. 
Among the six participants who were able to complete the interaction task under the \textsc{NVR} condition, five were still able to complete the task with \textsc{EnVisionVR}. However, P9 with little remaining vision was not able to complete the task with \textsc{EnVisionVR} as he felt the main objects indication function provided conflicting information by reporting an object directly behind him, which could not be confirmed easily using vision. Most participants of the subgroup who regularly use assistive technology (6 out of 8) were not able to complete the interaction task in the NVR condition, and \textsc{EnVisionVR} was able to support five out of these six participants to complete the interaction task. Overall, object interaction task completion results show a $83.3\%-50\%=33.3\%$ improvement in task success rate with \textsc{EVR} compared with \textsc{NVR}.
A McNemar's test did not indicate a significant difference ($\chi^2=2.67, p=.102$) between the \textsc{NVR} and \textsc{EVR} object interaction task completion status.
For the subgroup of participants who regularly use assistive technology, a McNemar's test indicated a significant difference ($\chi^2=5.0, p<.05$) between the \textsc{NVR} and \textsc{EVR} object interaction task completion status, but did not reveal a significant difference ($\chi^2=1.0, p=.317$) for participants who do not regularly use assistive technology.

\begin{figure}[ht!]
    \centering
    \hspace{-0.1in}
    \includegraphics[width=\linewidth]{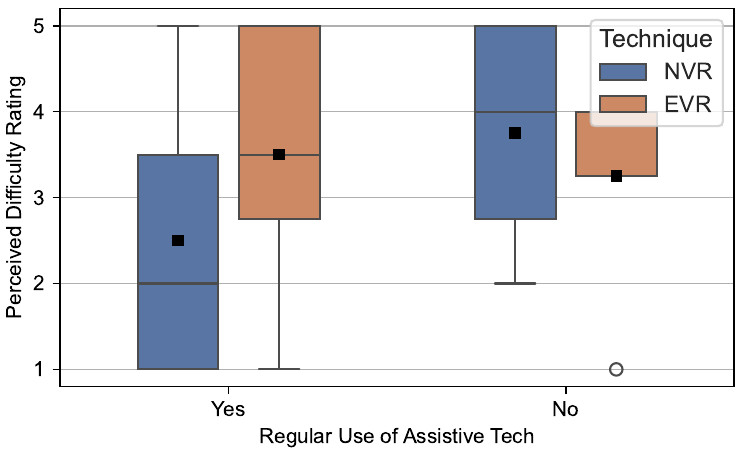}
    \caption[Perceived difficulty of the object interaction task.]{Object Interaction Task: Distribution of the perceived difficulty (higher score indicates lower perceived difficulty) for the \textsc{NVR} and \textsc{EVR} conditions for participants who regularly use assistive technology and for those who do not. Black squares indicate the mean value.}
    \label{fig:score-oi-difficulty-box}
\end{figure}

\Cref{fig:score-oi-difficulty-box} presents box plots of
the perceived difficulty of the task for the \textsc{NVR} ($M=2.92, SD=1.68$) and \textsc{EVR} ($M=3.42, SD=1.44$) conditions. 
Friedman tests did not reveal a significant difference ($\chi^2=2.78, p=.096$) between the \textsc{NVR} and \textsc{EVR} conditions for all participants, or for participants who do not regularly use assistive technology ($\chi^2<.01, p=1.0$), but revealed a significant difference for participants who regularly use assistive technology ($\chi^2=5.0, p<.05$).

\subsection{Interaction Behaviors}
\label{ssec:behaviors}

\begin{figure}[ht!]
    \centering
    \includegraphics[width=\linewidth]{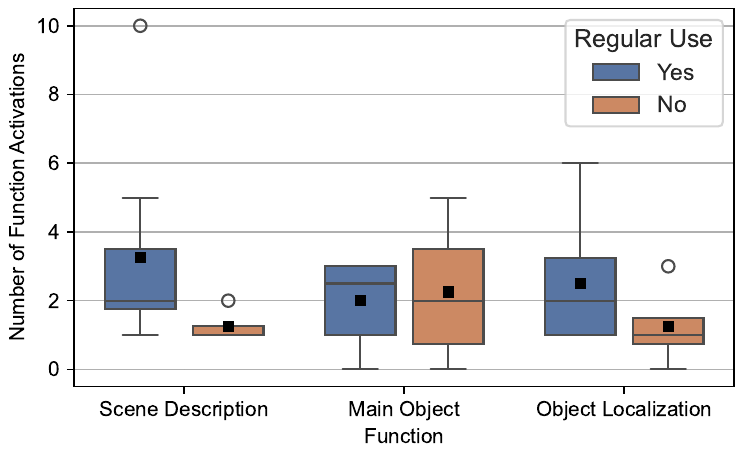}
    \caption[Distribution of the number of \textsc{EnVisionVR} function activations.]{Distribution of the number of \textsc{EnVisionVR} function activations for participants who regularly use assistive technology and for those who do not with labeled outliers. Black squares indicate the mean value.}
    \label{fig:function-activations}
\end{figure}

\Cref{fig:function-activations} plots the distribution of the number of \textsc{EnVisionVR} function activations for participants in the user study. The results show that the main objects indication function was activated a similar number of times
for participants who regularly use assistive technology ($M=2.00, SD=1.20$) and for participants who do not ($M=2.25, SD=2.22$). However, participants who regularly use assistive technology activated the scene description function more ($M=3.25, SD=3.01$) than participants who do not regularly use assistive technology ($M=1.25, SD=.500$). While a Mann-Whitney U test did not reveal a statistically significant difference ($U=6.50, p=.106$) in the number of scene description function activations, the rank-biserial correlation effect size was moderate ($r_{rb}=.594$), suggesting that there may be a practical difference between both groups. The object localization function was also activated more by participants who regularly use assistive technology ($M=2.50, SD=1.77$) compared with those who do not ($M=1.25, SD=1.26$), but the Mann-Whitney U test did not reveal a statistically significant difference ($U=8.50, p=.216, r_{rb}=.469$). This suggests that participants with less visual perception capability tend to rely more on high-level scene descriptions and the fine detail object localization function. Meanwhile, participants with different vision capabilities relied on the Main Objects Indication Function at a similar level.
    
\begin{figure}[ht!]
    \centering
    \includegraphics[width=\linewidth]{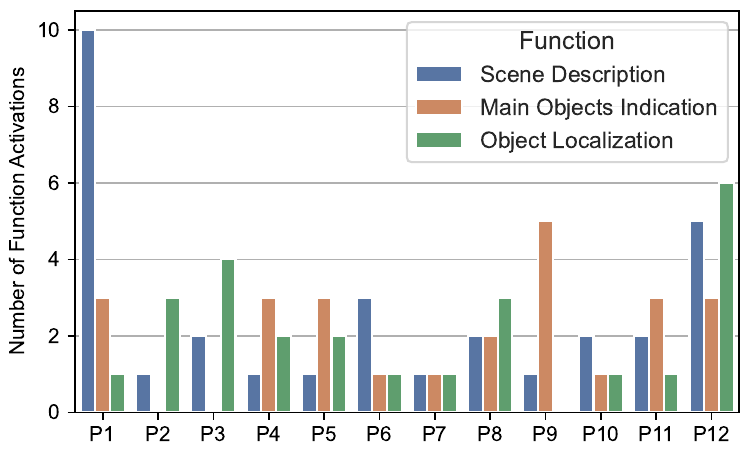}
    \caption[Total number of \textsc{EnVisionVR} function activations for each participant.]{Total number of \textsc{EnVisionVR} function activations for each participant for the scene understanding question, object localization task, and object interaction task. P2, P7, P9, and P11 do not regularly use assistive technology, while others do.}
    \label{fig:function-activations-individual}
\end{figure}

For participants who do not regularly use assistive technology, \textsc{EnVisionVR} appeared to complement their available vision. These participants used the scene description function less as they have enough residual vision to support their understanding of the scene, as evidenced by the performance of P2, P7, P9, and P11 in the scene understanding question in the \textsc{NVR} condition.
They were also able to precisely locate small virtual objects as evidenced by successful completion of the object interaction task under the \textsc{NVR} condition.
These participants used the main objects indication function more, likely because their residual vision does not allow them to explore a wide range in the scene, and they rely on the function to know what key objects are nearby.
Nevertheless, as \Cref{fig:function-activations-individual} indicates, the interaction behavior observations in \Cref{fig:function-activations} only describe general trends, and the usage of different functions for each participant can be vastly different. 

\subsection{Post-Study Interview}
\label{ssec:interview}

In the post-study interview, 11 out of 12 participants expressed a preference for \textsc{EnVisionVR} over the \textsc{NVR} condition. 
Key themes are summarized below.

\subsubsection{Level of Information Delivered.} 

The design of visual accessibility systems often faces a trade-off between the level of detail of information provided and the ease of use of the system. 
For the \textsc{EnVisionVR} system, participants liked how the scene description function was ``helpful to identify a new location'' (P2, P4, P5, P7, P10) with the ``correct amount of detail'' (P7) and ``gave you a picture of the scene'' (P3) and ``a general overview'' (P9, P10) to ``build an image up in your mind'' (P6). P11 commented that the tools could be more helpful if there was more detailed information. Overall, participants liked how the level of information delivered had an appropriate amount of detail to gain a sense of physical presence in the scene.

\begin{quote}
    \textit{``I felt that I was in a real bar or a restaurant and I'd gotten out my camera [to get a description using Be My Eyes], it was very very good.''} (P3)
\end{quote}

Participants also liked how the main objects indication function told the user about key objects nearby. 
They liked how it told the user ``if there is something on the right and left'' (P2) and ``helped users to be more confident by saying the names of things'' (P5), but found the spatial tone to indicate the object location to be redundant (P4), and it sometimes did not pick up objects near the user (P9).

Participants found the object localization function helpful in providing the precise location of individual objects. Participants found it helpful in ``telling [participants] how far [they] need to move'' (P2), locating the object and letting the user ``know the object is there'' (P5), and ``helping [participants] understand the direction of the object'' and ``gives good feedback'' (P8). P4 found the beeping helpful in telling whether the controller was getting closer to the object. Overall, participants found \textsc{EnVisionVR} helpful in delivering both high-level scene information and detailed object-level information.

\begin{quote}
\textit{``[With {EnVisionVR},] this is the first time that I have been able to do anything in VR. This is really promising, I think you’re on to something here.''} (P1)
\end{quote}


\subsubsection{System Latency.} 
Participants commented that existing vision accessibility systems they use in the physical world, such as Be My AI, take several seconds to return a description of the provided image. 
Participants contrasted this with their experience of \textsc{EnVisionVR}, which they found to be very responsive. P8 commented how she appreciated systems which give constant feedback on updates of what is in front of the user as the user moves.

\begin{quote}
    \textit{``The advantage here is that the answer comes instantly and doesn't take any time to process the question.''} (P3)
\end{quote}

\subsubsection{Consistency in System Design.} 
The user study revealed that inconsistency in the system design could pose usability barriers. For example, the command ``Where am I?'' for the scene description function describes the user's field of view, while the command ``What is near me?'' for the main objects indication function reads out the names of objects which are near the user but not necessarily in front of the user. The inconsistency in reference frames led to confusion among some participants (P1, P9).

\begin{quote}
    \textit{``Visually I could see where things were and I could move towards easily. The voice assistant wasn't giving me the instruction I needed, it didn't quite work. When I was looking for that `brew button', it said it was next to me but I couldn't see it."} (P9)
\end{quote}

As the scene descriptions were pre-baked based on images of the scene, 
there can be different names for the same object in the scene description function and the main objects indication function. For example, the bookholder on the desk in Anchor 1 of the test scene was referred to as a computer in scene descriptions, which can cause confusion for users.


\subsubsection{User Agency.} Participants also commented on different aspects of user agency over the system. These included user control on what information to deliver, the speed of delivery, and the level of detail of delivered information.

\begin{quote}
    \textit{``I'd like this experience to have a speedier read-out speed, close to 200\%. That really should be customisable. It'd be nice to have two levels of description, detailed and then summary.''} (P1)
\end{quote}

\begin{quote}
    \textit{``I'd like a mode where I could scan a room, turning in my chair, and keep hearing an updated description of what's in front of me."} (P1)
\end{quote}

\begin{quote}
    \textit{``For me maybe [the voice description] was a bit slow. If you are in a new environment you don't want it too fast.''} (P6)
\end{quote}

A number of participants also suggested that the system could support more voice commands to improve user agency. P7 commented that the system was helpful but required users to memorize the different speech commands for each function. P12 commented that she had difficulty in trying to remember how to phrase the question.


\begin{quote}
    \textit{``The only thing I would say is maybe broaden the wording used to launch the command... If it was a bit more open in terms of voice commands.''} (P6)
\end{quote}

\section{Design Implications}

Results from our evaluation study reveal important design implications for VLM-assisted interactive systems for visual accessibility design in VR.
Through these guidelines, we intend to assist designers and developers in creating more inclusive immersive systems for BLV users.

\begin{enumerate}
    \item \textbf{Different levels of vision require different levels of information.}
    \Cref{fig:function-activations} not only suggests that participants with less vision capability generally require more information than those with better vision. Most importantly, it highlights the uneven distribution of high-level scene information and detailed object-level information required by both participant subgroups, and how this uneven distribution is different for both groups. Participants with less vision capability rely primarily on visual accessibility systems to perceive visual information as compared to their residual vision, so they are inclined to ask for different levels of visual information. 
    Participants with better vision capability primarily used their residual vision and the vision accessibility tool to complement and verify what they saw. For example, P9 was confused when the system said that a `Brew Button' was close to him, but he could not see it in his view.
    \item \textbf{Information should be promptly conveyed to support interaction.} As BLV users rely on visual accessibility systems to support their understanding of 3D scenes, such systems need to deliver rich information with low latency. As P3 commented, \textsc{EnVisionVR} provided descriptions similar to those generated by human volunteers in Be My Eyes. P3 liked how \textsc{EnVisionVR} was able to generate descriptions instantly, unlike Be My Eyes which required users to wait for the human response or Be My AI which also required around 5 seconds to receive AI responses from the server. 
    \item \textbf{Spatial information should be expressed consistently.} 
    VLM-based accessibility systems should ensure that
    the delivered spatial information is consistent with the user's perception model. The system could deliver scene descriptions with respect to the user's field of view, with the option to keep providing an updated description of what is in front of the user (P1, P8), or deliver descriptions of the entire scene from a third person view, instead of only what is in front of the user (P9), but the reference frame has to be made explicit and intuitive for the user.
    Additionally, the delivered information should also be consistent among different functions. 
    The same applies to the consistency in the names of objects mentioned in different functions. 
    \item \textbf{New systems should be consistent with established assistive technology workflows.} For example, participants who regularly use screen readers such as P1 and P6 preferred the speed of the voice assistant to be faster and adjustable. However, participants who do not regularly use assistive technology did not express preferences in system verbosity or reading speed. 
    The same applies for customizable level of detail of descriptions to cater to different user preferences (P1, P6, P11).
    P1, P6, and P7 also suggested improving the number of supported speech commands, as it would otherwise pose a heavy mental load and cause unnecessary distractions.
    These observations suggest that the flexibility in speech commands and accuracy in recognizing user intent as seen in many speech-based accessibility systems is also crucial for VR visual accessibility systems to adapt to different users to reduce their effort and improve user agency.
    \item \textbf{Redundancy should be provided in system input and output.} Our user study demonstrated how \textsc{EnVisionVR} was helpful in assisting BLV users
    through a combination of audio cues, speech descriptions, and haptic vibrations.
    P2 and P8 appreciated how the haptic vibration together with audio/speech confirmation reassured them by indicating
    that the controller had reached the object of interest.
    Conversely, the lack of certain feedback modalities can take away user confidence, as in the case of P4 when she felt a vibration in the controller but received no speech feedback.
    This highlights the need to 
    include system input and output redundancy and 
    ensure that information from different input and output channels are consistent, which aligns with the design principles (DP1 and DP2)
    identified by Dudley et al.~\cite{dudley2023inclusive}.
\end{enumerate}

\section{Discussion}

\textsc{EnVisionVR} represents an original integration of high-level natural language scene descriptions and detailed object-level speech, audio, and haptic cues for object localization and interaction. We complement previous work on visual accessibility design in VR by incorporating VLMs to provide detailed scene descriptions to extend works such as SeeingVR~\citep{zhao2019seeingvr} and VRBubble~\citep{ji2022vrbubble} which convert visual information to speech and audio, while also following Canetroller~\citep{canetroller} and VIVR~\citep{kim2020vivr} in incorporating different feedback modalities to convey visual information such as the presence of a virtual object. We also demonstrate how it is possible to leverage speech, audio, and haptic information together to design a multimodal system for VR visual accessibility design. Results from the user study show good promise in terms of supporting BLV users to enjoy VR experiences with the greatest benefit seemingly afforded to blind users or users with less usable vision. 

The study results also reveal how \textsc{EnVisionVR} could be further improved.
As \textsc{EnVisionVR} is intended to provide a proof-of-concept of how VLMs can be applied with other interaction modalities for visual accessibility design for VR content, the scene descriptions are pre-baked. This limits the current approach to static VR scenes.
Future design iterations will aim to provide scene descriptions for dynamic VR scenes.

The evaluative study found different participants had different preferences in the verbosity and level of detail of scene descriptions. 
These examples demonstrate the significance of incorporating the ability to customize features for individual preferences, as well as adaptations for each user as they become more accustomed to the system.
Additionally, we acknowledge that \textsc{EnVisionVR} is primarily a speech-driven interface with a limited number of supported commands. Future design iterations of \textsc{EnVisionVR} will 
allow users to access more object and scene-level information through alternative and complementary forms of interaction. 


\section{Conclusion}

This paper presents \textsc{EnVisionVR}, a proof-of-concept visual accessibility tool for VR based on scene descriptions and object-level guidance powered by VLMs, speech and audio cues, and haptic feedback. 
Our evaluation study with 12 BLV participants demonstrates the effectiveness of \textsc{EnVisionVR} in assisting scene understanding,
object localization (41.7\% increase in task success rate), and object interaction (33.3\% increase in task success rate) for BLV users compared with the condition without visual accessibility features. We also summarize a list of design implications covering five different aspects of visual accessibility. 
We hope these findings and contributions will advance research in this space and ultimately lead to more inclusive VR experiences.

\section*{Supplemental Material}
The online appendix is available at \url{https://osf.io/zb2ak/}.

\section*{Acknowledgments}
Junlong Chen is supported by the China Scholarship Council and Cambridge Trust.
This work is also supported by the Engineering and Physical Sciences Research Council (EPSRC), through the following grants: \textit{Inclusive Immersion: Inclusive Design of Immersive Content} (EP/S027637/1 and EP/S027432/1) and \textit{Towards an Equitable Social VR} (EP/W025698/1 and EP/W02456X/1). The authors thank Open Inclusion for their help in recruiting the research participants and administering the user research.            

\bibliographystyle{plainnat}
\bibliography{main}

\begin{thebibliography}{33}
\providecommand{\natexlab}[1]{#1}
\providecommand{\url}[1]{\texttt{#1}}
\expandafter\ifx\csname urlstyle\endcsname\relax
  \providecommand{\doi}[1]{doi: #1}\else
  \providecommand{\doi}{doi: \begingroup \urlstyle{rm}\Url}\fi

\bibitem[Bongini et~al.(2023)Bongini, Becattini, and Del~Bimbo]{bongini2023gpt}
Pietro Bongini, Federico Becattini, and Alberto Del~Bimbo.
\newblock {Is GPT-3 All You Need for Visual Question Answering in Cultural Heritage?}
\newblock In \emph{Computer Vision--ECCV 2022 Workshops: Tel Aviv, Israel, October 23--27, 2022, Proceedings, Part I}, pages 268--281. Springer, 2023.

\bibitem[Borodin et~al.(2010)Borodin, Bigham, Dausch, and Ramakrishnan]{borodin2010more}
Yevgen Borodin, Jeffrey~P Bigham, Glenn Dausch, and IV~Ramakrishnan.
\newblock {More than meets the eye: A survey of screen-reader browsing strategies}.
\newblock In \emph{Proceedings of the 2010 International Cross Disciplinary Conference on Web Accessibility (W4A)}, pages 1--10, 2010.

\bibitem[Cho et~al.(2021)Cho, Lei, Tan, and Bansal]{pmlr-v139-cho21a}
Jaemin Cho, Jie Lei, Hao Tan, and Mohit Bansal.
\newblock {Unifying Vision-and-Language Tasks via Text Generation}.
\newblock In Marina Meila and Tong Zhang, editors, \emph{Proceedings of the 38th International Conference on Machine Learning}, volume 139 of \emph{Proceedings of Machine Learning Research}, pages 1931--1942. PMLR, 18--24 Jul 2021.

\bibitem[Ciccone et~al.(2023)Ciccone, Bailey, and Lewis]{ciccone2023next}
Brendan~A Ciccone, Shannon~KT Bailey, and Joanna~E Lewis.
\newblock The next generation of virtual reality: recommendations for accessible and ergonomic design.
\newblock \emph{Ergonomics in Design}, 31\penalty0 (2):\penalty0 24--27, 2023.

\bibitem[Creed et~al.(2024)Creed, Al-Kalbani, Theil, Sarcar, and Williams]{creed2024inclusive}
Chris Creed, Maadh Al-Kalbani, Arthur Theil, Sayan Sarcar, and Ian Williams.
\newblock {Inclusive AR/VR: accessibility barriers for immersive technologies}.
\newblock \emph{Universal Access in the Information Society}, 23\penalty0 (1):\penalty0 59--73, 2024.

\bibitem[Dang et~al.(2023)Dang, Korreshi, Iqbal, and Lee]{dang2023opportunities}
Khang Dang, Hamdi Korreshi, Yasir Iqbal, and Sooyeon Lee.
\newblock {Opportunities for Accessible Virtual Reality Design for Immersive Musical Performances for Blind and Low-Vision People}.
\newblock In \emph{Proceedings of the 2023 ACM Symposium on Spatial User Interaction}, pages 1--21, 2023.

\bibitem[De~La~Torre et~al.(2024)De~La~Torre, Fang, Huang, Banburski-Fahey, Amores~Fernandez, and Lanier]{de2024llmr}
Fernanda De~La~Torre, Cathy~Mengying Fang, Han Huang, Andrzej Banburski-Fahey, Judith Amores~Fernandez, and Jaron Lanier.
\newblock {LLMR: Real-time prompting of interactive worlds using large language models}.
\newblock In \emph{Proceedings of the CHI Conference on Human Factors in Computing Systems}, pages 1--22, 2024.

\bibitem[Di~Blas et~al.(2004)Di~Blas, Paolini, Speroni, et~al.]{di2004usable}
Nicoletta Di~Blas, Paolo Paolini, Marco Speroni, et~al.
\newblock {“Usable Accessibility” to the Web for Blind Users}.
\newblock In \emph{Proceedings of 8th ERCIM Workshop: User Interfaces for All, Vienna}, 2004.

\bibitem[Dudley et~al.(2023)Dudley, Yin, Garaj, and Kristensson]{dudley2023inclusive}
John Dudley, Lulu Yin, Vanja Garaj, and Per~Ola Kristensson.
\newblock {Inclusive Immersion: a review of efforts to improve accessibility in virtual reality, augmented reality and the metaverse}.
\newblock \emph{Virtual Reality}, 27\penalty0 (4):\penalty0 2989--3020, 2023.

\bibitem[Eyes(2023)]{BeMyAI}
Be~My Eyes, Sep 2023.
\newblock Available at: \url{https://www.bemyeyes.com/blog/announcing-be-my-ai}. Accessed on December 4th 2024.

\bibitem[Gonzalez~Penuela et~al.(2024)Gonzalez~Penuela, Collins, Bennett, and Azenkot]{gonzalez2024investigating}
Ricardo~E Gonzalez~Penuela, Jazmin Collins, Cynthia Bennett, and Shiri Azenkot.
\newblock {Investigating Use Cases of AI-Powered Scene Description Applications for Blind and Low Vision People}.
\newblock In \emph{Proceedings of the CHI Conference on Human Factors in Computing Systems}, pages 1--21, 2024.

\bibitem[Herskovitz et~al.(2020)Herskovitz, Wu, White, Pavel, Reyes, Guo, and Bigham]{herskovitz2020making}
Jaylin Herskovitz, Jason Wu, Samuel White, Amy Pavel, Gabriel Reyes, Anhong Guo, and Jeffrey~P Bigham.
\newblock {Making Mobile Augmented Reality Applications Accessible}.
\newblock In \emph{Proceedings of the 22nd International ACM SIGACCESS Conference on Computers and Accessibility}, pages 1--14, 2020.

\bibitem[Inclusion()]{OpenInclusion}
Open Inclusion.
\newblock {Home - Open Inclusion}.
\newblock Available at: https://openinclusion.com/. Accessed on Jan. 19th, 2025.

\bibitem[Ji et~al.(2022)Ji, Cochran, and Zhao]{ji2022vrbubble}
Tiger~F Ji, Brianna Cochran, and Yuhang Zhao.
\newblock {VRBubble: Enhancing peripheral awareness of avatars for people with visual impairments in social virtual reality}.
\newblock In \emph{Proceedings of the 24th International ACM SIGACCESS Conference on Computers and Accessibility}, pages 1--17, 2022.

\bibitem[Jiang et~al.(2023)Jiang, Phutane, and Azenkot]{jiang2023beyond}
Lucy Jiang, Mahika Phutane, and Shiri Azenkot.
\newblock {Beyond Audio Description: Exploring 360° Video Accessibility with Blind and Low Vision Users Through Collaborative Creation}.
\newblock In \emph{Proceedings of the 25th international ACM SIGACCESS conference on computers and accessibility}, pages 1--17, 2023.

\bibitem[Kearney-Volpe and Hurst(2021)]{accessibleWebDev2021}
Claire Kearney-Volpe and Amy Hurst.
\newblock {Accessible Web Development: Opportunities to Improve the Education and Practice of Web Development with a Screen Reader}.
\newblock \emph{ACM Trans. Access. Comput.}, 14\penalty0 (2), jul 2021.
\newblock ISSN 1936-7228.
\newblock \doi{10.1145/3458024}.

\bibitem[Kim(2020)]{kim2020vivr}
Jinmo Kim.
\newblock {VIVR: Presence of immersive interaction for visual impairment virtual reality}.
\newblock \emph{IEEE Access}, 8:\penalty0 196151--196159, 2020.

\bibitem[Luo et~al.(2022)Luo, Xi, Zhang, and Ma]{luo2022vc}
Ziyang Luo, Yadong Xi, Rongsheng Zhang, and Jing Ma.
\newblock {VC-GPT: Visual Conditioned GPT for End-to-End Generative Vision-and-Language Pre-training}.
\newblock \emph{arXiv preprint arXiv:2201.12723}, 2022.

\bibitem[Masnadi et~al.(2020)Masnadi, Williamson, Gonz{\'a}lez, and LaViola]{masnadi2020vriassist}
Sina Masnadi, Brian Williamson, Andr{\'e}s N~Vargas Gonz{\'a}lez, and Joseph~J LaViola.
\newblock {VRiAssist: An eye-tracked virtual reality low vision assistance tool}.
\newblock In \emph{2020 IEEE Conference on Virtual Reality and 3D User Interfaces Abstracts and Workshops (VRW)}, pages 808--809. IEEE, 2020.

\bibitem[Microsoft(2021)]{microsoft-seeing-ai}
Microsoft.
\newblock {Seeing AI}.
\newblock \url{https://www.microsoft.com/en-us/ai/seeing-ai}, September 2021.

\bibitem[Morris et~al.(2018)Morris, Johnson, Bennett, and Cutrell]{altText2018}
Meredith~Ringel Morris, Jazette Johnson, Cynthia~L. Bennett, and Edward Cutrell.
\newblock {Rich Representations of Visual Content for Screen Reader Users}.
\newblock In \emph{Proceedings of the 2018 CHI Conference on Human Factors in Computing Systems}, CHI '18, page 1–11, New York, NY, USA, 2018. Association for Computing Machinery.
\newblock ISBN 9781450356206.
\newblock \doi{10.1145/3173574.3173633}.

\bibitem[Naikar et~al.(2024)Naikar, Subramanian, and Tigwell]{naikar2024accessibility}
Vinaya~Hanumant Naikar, Shwetha Subramanian, and Garreth~W Tigwell.
\newblock {Accessibility Feature Implementation Within Free VR Experiences}.
\newblock In \emph{Extended Abstracts of the CHI Conference on Human Factors in Computing Systems}, pages 1--9, 2024.

\bibitem[Salin et~al.(2022)Salin, Farah, Ayache, and Favre]{salin2022vision}
Emmanuelle Salin, Badreddine Farah, St{\'e}phane Ayache, and Benoit Favre.
\newblock {Are Vision-Language Transformers Learning Multimodal Representations? A Probing Perspective}.
\newblock In \emph{Proceedings of the AAAI Conference on Artificial Intelligence}, volume~36, pages 11248--11257, 2022.

\bibitem[Southwell and Slater(2013)]{southwell2013evaluation}
Kristina~L Southwell and Jacquelyn Slater.
\newblock {An Evaluation of Finding Aid Accessibility for Screen Readers}.
\newblock \emph{Information Technology and Libraries}, 32\penalty0 (3):\penalty0 34--46, 2013.

\bibitem[Te{\'o}filo et~al.(2018)Te{\'o}filo, Lucena, Nascimento, Miyagawa, and Maciel]{teofilo2018evaluating}
Mauro Te{\'o}filo, Vicente~F Lucena, Josiane Nascimento, Taynah Miyagawa, and Francimar Maciel.
\newblock Evaluating accessibility features designed for virtual reality context.
\newblock In \emph{2018 IEEE international conference on consumer electronics (ICCE)}, pages 1--6. IEEE, 2018.

\bibitem[Unity(2021)]{vr-escape-room}
Unity.
\newblock {VR Beginner: The Escape Room}.
\newblock \url{https://assetstore.unity.com/packages/templates/tutorials/vr-beginner-the-escape-room-163264}, October 2021.

\bibitem[{WebAim}(2024)]{WebAim-10}
{WebAim}.
\newblock {{Screen Reader User Survey \#10 Results}}.
\newblock \url{https://webaim.org/projects/screenreadersurvey10/}, 2024.
\newblock [Online; accessed 13-August-2024].

\bibitem[Williams et~al.(2019)Williams, Clarke, Gardiner, Zimmerman, and Tomasic]{williams2019find}
Kristin Williams, Taylor Clarke, Steve Gardiner, John Zimmerman, and Anthony Tomasic.
\newblock {Find and Seek: Assessing the Impact of Table Navigation on Information Look-up with a Screen Reader}.
\newblock \emph{ACM Transactions on Accessible Computing (TACCESS)}, 12:\penalty0 1--23, 2019.

\bibitem[Wu et~al.(2017)Wu, Wieland, Farivar, and Schiller]{alttext2017}
Shaomei Wu, Jeffrey Wieland, Omid Farivar, and Julie Schiller.
\newblock {Automatic Alt-Text: Computer-Generated Image Descriptions for Blind Users on a Social Network Service}.
\newblock In \emph{Proceedings of the 2017 ACM Conference on Computer Supported Cooperative Work and Social Computing}, CSCW '17, page 1180–1192, New York, NY, USA, 2017. Association for Computing Machinery.
\newblock ISBN 9781450343350.
\newblock \doi{10.1145/2998181.2998364}.

\bibitem[Zhang and Chen(2023)]{zhang2023gpt4mia}
Yizhe Zhang and Danny~Z Chen.
\newblock {GPT4MIA: Utilizing Geneative Pre-trained Transformer (GPT-3) as A Plug-and-Play Transductive Model for Medical Image Analysis}.
\newblock \emph{arXiv preprint arXiv:2302.08722}, 2023.

\bibitem[Zhao et~al.(2018)Zhao, Bennett, Benko, Cutrell, Holz, Morris, and Sinclair]{canetroller}
Yuhang Zhao, Cynthia~L. Bennett, Hrvoje Benko, Edward Cutrell, Christian Holz, Meredith~Ringel Morris, and Mike Sinclair.
\newblock {Enabling People with Visual Impairments to Navigate Virtual Reality with a Haptic and Auditory Cane Simulation}.
\newblock In \emph{Proceedings of the 2018 CHI Conference on Human Factors in Computing Systems}, CHI '18, page 1–14, New York, NY, USA, 2018. Association for Computing Machinery.
\newblock ISBN 9781450356206.
\newblock \doi{10.1145/3173574.3173690}.
\newblock URL \url{https://doi.org/10.1145/3173574.3173690}.

\bibitem[Zhao et~al.(2019)Zhao, Cutrell, Holz, Morris, Ofek, and Wilson]{zhao2019seeingvr}
Yuhang Zhao, Edward Cutrell, Christian Holz, Meredith~Ringel Morris, Eyal Ofek, and Andrew~D Wilson.
\newblock {SeeingVR: A set of tools to make virtual reality more accessible to people with low vision}.
\newblock In \emph{Proceedings of the 2019 CHI conference on human factors in computing systems}, pages 1--14, 2019.

\bibitem[Zong et~al.(2022)Zong, Lee, Lundgard, Jang, Hajas, and Satyanarayan]{zong2022rich}
Jonathan Zong, Crystal Lee, Alan Lundgard, JiWoong Jang, Daniel Hajas, and Arvind Satyanarayan.
\newblock {Rich Screen Reader Experiences for Accessible Data Visualization}.
\newblock In \emph{Computer Graphics Forum}, volume~41, pages 15--27. Wiley Online Library, 2022.

\end{thebibliography}

\begin{IEEEbiography}
[{\includegraphics[width=1in,height=1.25in,clip,keepaspectratio]{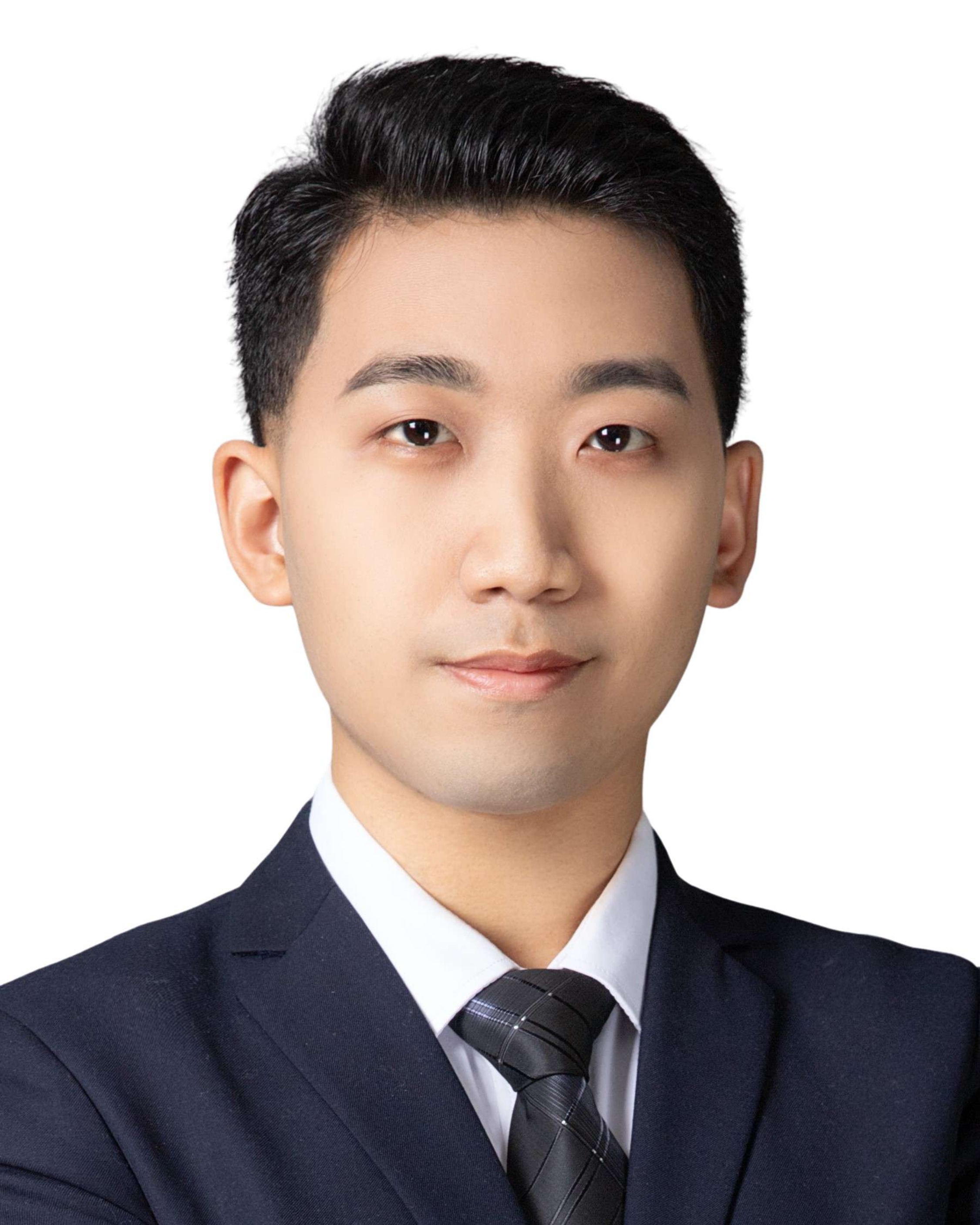}}]{Junlong Chen} is a PhD Student at the Department of Engineering, University of Cambridge. His research interests include scene interpretation for intelligent multimodal interactive systems and accessibility design.
\end{IEEEbiography}
\vspace{-33pt}
\begin{IEEEbiography}
[{\includegraphics[width=1in,height=1.25in,clip,keepaspectratio]{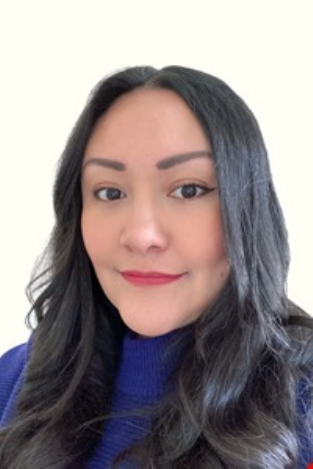}}]{Rosella P. Galindo Esparza} is a Research Fellow at the Brunel Design School, Brunel University of London. She is a member of the Brunel Digital Design Lab, and her research focuses on human-computer interaction design for immersive technologies, accessibility, and social inclusion.
\end{IEEEbiography}
\vspace{-33pt}
\begin{IEEEbiography}
[{\includegraphics[width=1in,height=1.25in,clip,keepaspectratio]{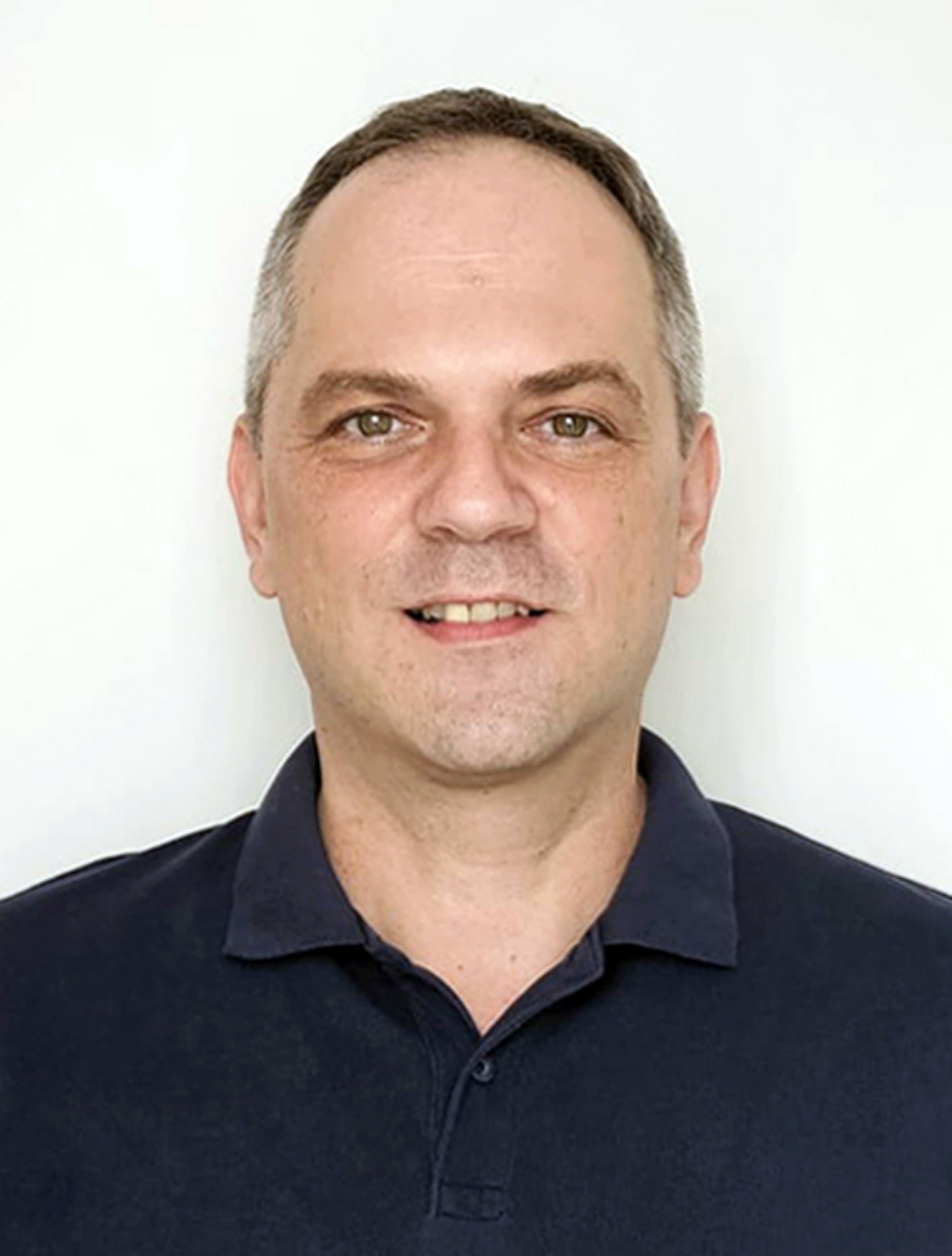}}]{Vanja Garaj} is a Professor of Design and the Director of Research at the Brunel Design School, Brunel University of London, where he also leads the Brunel Digital Design Lab, a research group specialising in design-led technology innovation. His research interests include human-computer interaction, accesibility and inclusive design.
\end{IEEEbiography}
\vspace{-33pt}
\begin{IEEEbiography}
[{\includegraphics[width=1in,height=1.25in,clip,keepaspectratio]{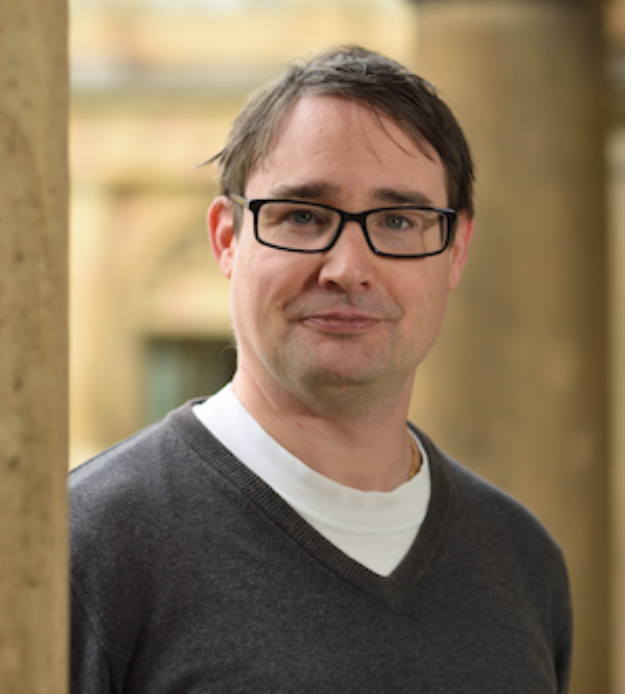}}]{Per Ola Kristensson} is a Professor of Interactive Systems Engineering at the Department of Engineering, University of Cambridge and a Fellow of Trinity College, Cambridge. He is a co-founder and co-director of the Centre for Human-Inspired Artificial Intelligence (CHIA) at the University of Cambridge.
\end{IEEEbiography}
\vspace{-33pt}
\begin{IEEEbiography}
[{\includegraphics[width=1in,height=1.25in,clip,keepaspectratio]{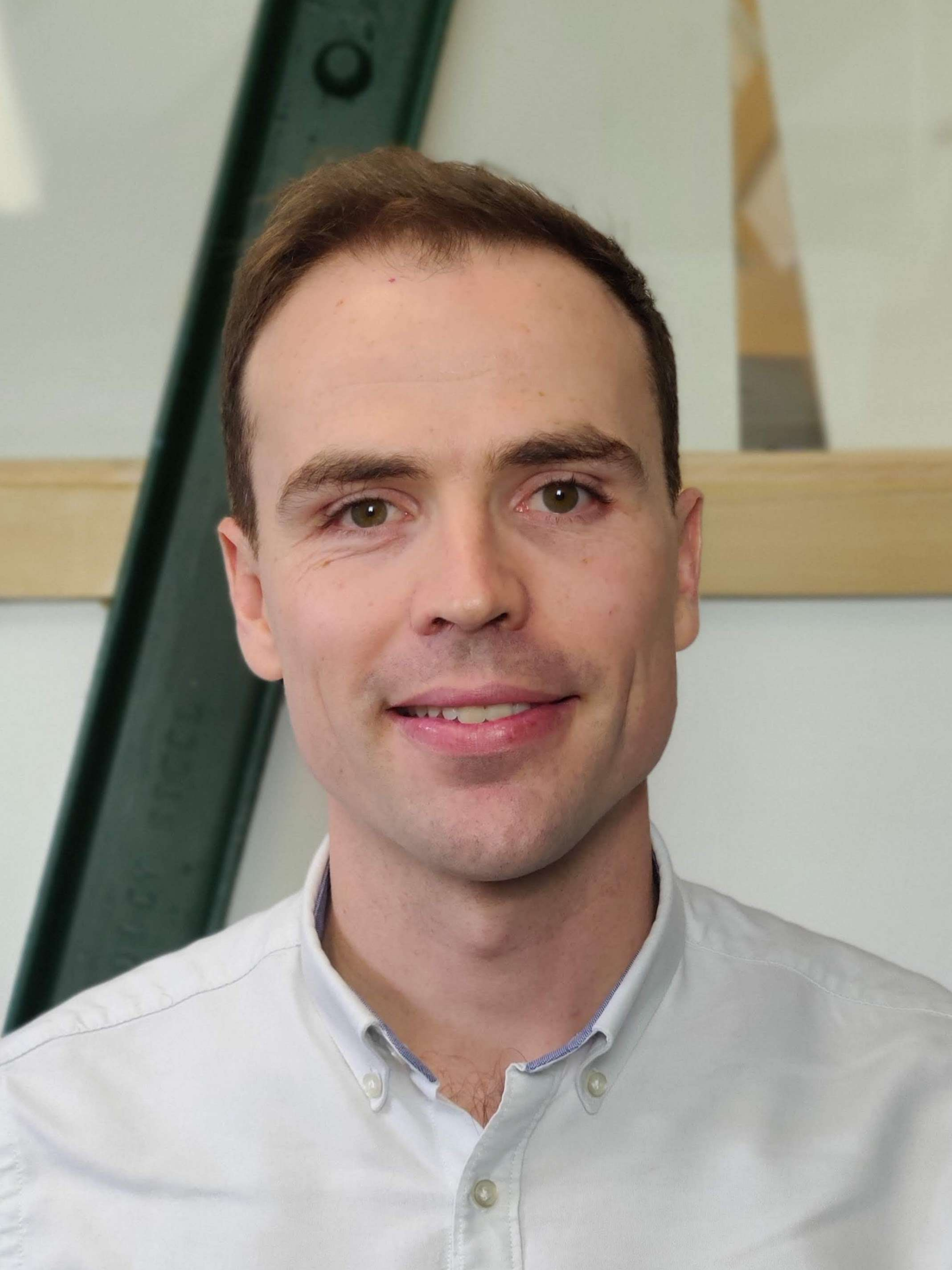}}]{John J. Dudley} is an Associate Teaching Professor of Machine Learning and Machine Intelligence at the Department of Engineering, University of Cambridge. He is a member of the Computational and Biological Learning Lab and his research focusses on the design of interactive systems that dynamically adapt to user needs and behaviours.
\end{IEEEbiography}



\end{document}